\documentclass[preprint,12pt]{elsarticle}

\usepackage{framed,multirow}

\usepackage{amssymb}
\usepackage{latexsym}
\usepackage{amsmath} 

\usepackage{url}
\usepackage{xcolor}

\usepackage{hyperref}
\usepackage{algorithm,algpseudocode}
\algnewcommand{\LeftComment}[1]{\Statex \(\triangleright\) #1}
\definecolor{newcolor}{rgb}{.8,.349,.1}

\journal{}

\begin{document}


\begin{frontmatter}

\title{NucleiMix: Realistic Data Augmentation for Nuclei Instance Segmentation}%

\author[1]{Jiamu Wang}
\author[1]{Jin Tae Kwak\corref{cor1}}
\cortext[cor1]{Corresponding author: 
  Tel.: +82-2-3290-3241;  
  Email: jkwak@korea.ac.kr;}

\address[1]{School of Electrical Engineering, Korea University, Seoul 02841, Republic of Korea}


\begin{abstract}
Nuclei instance segmentation is an essential task in pathology image analysis, serving as the foundation for many downstream applications. The release of several public datasets has significantly advanced research in this area, yet many existing methods struggle with data imbalance issues. To address this challenge, this study introduces a data augmentation method, called as NucleiMix, that is designed to balance the distribution of nuclei types by increasing the number of rare-type nuclei within datasets. NucleiMix operates in two phases. In the first phase, it identifies candidate locations similar to the surroundings of rare-type nuclei and inserts rare-type nuclei into the candidate locations. In the second phase, it employs a progressive inpainting strategy using a pre-trained diffusion model to seamlessly integrate rare-type nuclei into their new environments in replacement of major-type nuclei or background locations. We systematically evaluate the effectiveness of NucleiMix on three public datasets using two popular nuclei instance segmentation models. The results demonstrate the superior ability of NucleiMix to synthesize realistic rare-type nuclei and to enhance the quality of nuclei segmentation and classification in an accurate and robust manner.
\end{abstract}


\end{frontmatter}


\section{Introduction}

In recent years, the combination of large-scale pathology data and advanced artificial intelligence techniques has fueled the development and advancement of numerous digital pathology tools. These include cancer detection and grading~\cite{cancer}, tissue sub-typing~\cite{subtype}, nuclei segmentation and classification~\cite{consep}, and survival prediction~\cite{survive}. Among these, nuclei instance segmentation has gained much attention due to its critical role in pathology image analysis, as the count, shape, and distribution of nuclei are tightly related to the nature of diseases and cellular microenvironments~\cite{nuseg}. 
Nuclei instance segmentation involves identifying and delineating individual cell nuclei within pathology images. Deep learning models have substantially improved the accuracy of nuclei instance segmentation ~\cite{consep,glysac}, but they generally require a large amount of labeled datasets. Unfortunately, acquiring these labels is both costly and time-consuming, presenting a significant challenge.
Despite such challenges, there arise several public datasets, containing extensive annotations with differing types of cells and nuclei, such as ConSeP~\cite{consep}, MoNuSAC~\cite{monusac}, GLySAC~\cite{glysac}, and PanNuke~\cite{pannuke}. These datasets have formed the basis for the development and validation of various nuclei segmentation and classification models~\cite{dataset}. However, the types of cells/nuclei and their distributions substantially vary, i.e., such a dataset, in general, is greatly imbalanced. This imbalance, particularly the under-representation of rare cells/nuclei types, substantially contributes to the performance degradation of the deep learning models~\cite{imbalance}. Data augmentation techniques have proved efficient and effective in addressing the shortcomings of existing data and annotations by enlarging the dataset size and enriching data diversity. Earlier methods including random flipping, rotating, and cropping are popular and have been successfully applied to numerous applications [\cite{crop}, \cite{rotate}, \cite{rotate1}, \cite{flip}]. Later, more sophisticated mix-based data augmentation techniques, such as MixUp~\citep{mixup}, CutOut~\citep{cutout}, CutMix~\citep{cutmix}, CopyPaste~\cite{copypaste}, and GradMix~\citep{gradmix}, have been proposed to further enhance the performance of deep learning models. 
These existing mix-based data augmentation methods often overlook the effect of augmentation on neighboring regions, leading to unnecessary artifacts, particularly in nuclei data. For example, CutMix crops a whole bounding box including an object, e.g., a nucleus, and pastes it into a random location, resulting in an obvious rectangular artifact. CopyPaste crops only the object and paste it randomly. GradMix proposes to conduct a weighted sum between the cropped region and the background region but still leaves an unnatural transition zone around the object.

To address the mentioned challenges, we propose NucleiMix, a method designed to create realistic augmented images with various instances of nuclei. NucleiMix aims to balance and enrich nuclei data, thereby leading to improved nuclei instance segmentation. NucleiMix adopts a copy-paste approach to increase the number of under-represented instances of nuclei by using the existing instances. In other words, it copies an under-represented nucleus from the dataset and pastes it into a target location. However, a simple copy-paste approach can introduce the following issues: 
1) Random location: it randomly chooses the target regions that may severely differ from the original environment of the selected nuclei; 
2) Artifacts: it often creates unrealistic and unnecessary artifacts around the pasted nuclei;
3) Obscure boundary: it often loses the clarity of the nuclei boundary. 
To tackle these issues, NucleiMix introduces two strategies: 1) Context-aware probabilistic sampling and 2) Progressive inpainting using a diffusion model~\citep{sde, dm, ddpm, mcg}. 
Context-aware probabilistic sampling employs an unsupervised Gaussian Mixture Model (GMM)~\citep{gmm} to learn the distribution of nuclei along with their surrounding regions and sample target nuclei with environments resembling those of under-represented nuclei. For each target region, it samples the most likely nucleus from the pool of under-represented nuclei.
Progressive inpainting with a diffusion model generates realistic backgrounds for the under-represented nuclei in two steps, ensuring seamless integration with the existing environment of the target region. In the first step, it removes the nucleus in the target region and then inpaints it to create a smooth, clean background. In the second step, it places the under-represented nucleus along with its immediate one-layer contour (preserving its immediate background information) and removes its second and third contours, establishing a transitional zone. This transitional zone is inpainted to smoothly blend the nucleus with the surrounding area.

To prove the effectiveness of our method, we conduct extensive experiments on three public datasets, including CoNSeP\citep{consep}, GLySAC~\citep{glysac}, and MoNuSAC~\citep{monusac}. The experimental results show that NucleiMix is able to generate realistic, augmented images for nuclei data and mitigate the challenges posed by data imbalance, leading to improved nuclei instance segmentation.

\section{Related Works}
Enhancing dataset size and diversity significantly improves the robustness of deep learning models and helps mitigate overfitting. However, collecting medical image datasets is notably challenging and costly due to the exhaustive annotation required and concerns about patient privacy. In response to these challenges, data augmentation has emerged as an efficient solution to address the scarcity of data by effectively increasing dataset heterogeneity. NucleiMix, our novel mix-based data augmentation technique, is tailored specifically for histopathology datasets with accurate pixel-wise annotations. This section discusses traditional data augmentation methods alongside several innovative mix-based techniques.

\subsection{Traditional Data Augmentation}
Traditional data augmentation methods primarily focus on spatial and color transformations to diversify image presentations~\cite{survey}. Several spatial transformations have been widely adopted for various applications. For instance, random cropping randomly chooses a rectangular area within an image and uses it as a new image, helping the model extract features with different scales of views; random rotation randomly rotates an image by an angle within a specific range; vertical and horizontal flip apply vertical or horizontal mirroring to an image. Adjusting color and contrast is also a popular technique for data augmentation, which modifies the color tone or intensity levels of an image. For example, brightness adjustment alters the intensity values of an image, making it either darker or brighter compared to the original brightness; contrast adjustment applies a linear transformation to the intensity range~\cite{brightandcontrast} of an image; saturation modifies color intensity by multiplying a random color channel with a scalar factor~\cite{saturation}; hue adjustment randomly shifts color channels. Furthermore, there are other popular methods such as noise-based methods, adding Gaussian or Poisson noise on an image, and blur/unsharp-based methods, blurring images with Gaussian or median filters.

\subsection{Mix-based Data Augmentation}
Mix-based data augmentation methods typically combine two or more images and blend them together to create a composite image. MixUp~\cite{mixup} combines two random images by blending them in the alpha channel and generates a new label by taking a linear combination of the two original labels. CutOut~\cite{cutout} randomly masks a rectangular portion of an image, simulating dropout effects on input to deep learning models. Similar to CutOut, CutMix~\cite{cutmix} also masks a random portion of an image, but it fills the masked area with a portion taken from another image, simulating an occlusion in an image and increasing the diversity of image data. While these methods are generic and applicable to various tasks, there have also been specific efforts to develop tailored data augmentation methods for nuclei segmentation in pathology images. For instance, GradMix~\cite{gradmix} finds pairs of major-type nuclei and rare-type (under-represented) nuclei that meet specific shape constraints and replace the larger major nuclei with the relatively smaller rare-type nuclei. To alleviate the differences in the surrounding environments, GradMix linearly blends the neighboring region of the major-type nuclei with that of rare-type nuclei. Though it has shown to be effective in improving nuclei segmentation and classification, it has been only evaluated on a single nuclei dataset.

\section{Methodology}
Suppose that we are given a dataset of pathology images $\mathcal{D}$ containing nucleus-label pairs $\{(\alpha_i, y_i)\}_{i=1}^{N}$ where $\alpha_i$ is the $i$th nucleus, $y_i \in \{1,...,C\}$ is its corresponding class label, $N$ is the total number of nuclei, and $C$ is the cardinality of nuclear types. Among $C$ nuclear types, at least one type is under-represented in $\mathcal{D}$, identified as rare-type nuclei $\mathcal{A}^{r} = \{\alpha_i^r | i=1,..., N_r\}$, and the remaining types are regarded as major-type nuclei $\mathcal{A}^{m} = \{\alpha_i^m | i=1,..., N_m\}$ where $N_r$ and $N_m$ denote the number of rare-type and major-type nuclei, respectively, and $N_r << N_m$. In addition, certain locations $\mathcal{D}$ are collected and referred to as background locations $\mathcal{A}^{b} = \{\alpha_i^b | i=1,...,N_b\}$ where $N_b$ is the number of background locations. These locations contain neither major-type nor rare-type nuclei.
Given these three types of locations, NucleiMix strategically positions existing rare-type nuclei into locations where major-type nuclei are present or no nuclei are present (background locations), but where the neighboring environments are similar to those of rare-type nuclei. Then, it blends the original environment (of either major-type nuclei or background locations) with the inserted rare-type nuclei in a way that preserves the texture of the original environment as well as the shape and sharpness of the rare-type nuclei.
The detailed procedures are outlined in the Algorithm~\ref{alg}.

\subsection{Context-aware Probabilistic Sampling: Candidate Locations}

We consider three distinct types of locations: rare-type, major-type, and background locations. These are characterized by the presence and type of nuclei in the corresponding locations. For an image $I\in R^{H\times W}$, we randomly pick background locations for $\gamma H\times W$ times and we set $\gamma$ as $5e^{-4}$.
To analyze the neighboring environment of these locations, we first crop an image patch of size 224$\times$224 pixels around each identified location. Subsequently, we remove the central portion of size 112$\times$112 pixels for rare-type and major-type nuclei and fill the void region using a Navier-Stokes-based inpainting method~\cite{nsv}. 
Formally, we define rare-type patches as $\mathcal{P}^{r}$ = \{$\phi_i^r | i=1, ..., N_r$\}, major-type patches as $\mathcal{P}^{m}$ = \{$\phi_i^m | i=1, ..., N_m$\}, and background patches as $\mathcal{P}^{b}$ = \{$\phi_i^b | i=1, ..., N_b$\}. Here, $\phi_i^r$, $\phi_i^m$, and $\phi_i^b$ represent the patches centered around $\alpha_i^r$, $\alpha_i^m$, and $\alpha_i^b$, respectively. 

We use a pre-trained CTransPath~\cite{ctranspath} to extract a 768-dimensional feature for each patch, combined with a 4-dimensional Grey Level Co-occurrence Matrix (GLCM)~\cite{glcm} feature computed from a standard implementation of SciPy~\cite{scipy}. For each patch $\phi_i^r$, $\phi_i^m$, and $\phi_i^b$, it outputs feature vector $x_i^r, x_i^m, x_i^b \in \mathbb{R}^p$, where $p$ is 772, and PCA is used to reduce the dimension to $p=16$. This process results in three sets of feature vectors: rare-type feature vectors $\mathcal{X}^r=\{x_i^r\in \mathbb{R}^p, i=1, ..., N_r\}$, major-type feature vectors $\mathcal{X}^m=\{x_i^m\in \mathbb{R}^p, i=1, ..., N_m\}$, and background feature vectors $\mathcal{X}^b=\{x_i^b\in \mathbb{R}^p, i=1, ..., N_b\}$. We use a pre-trained CTransPath~\cite{ctranspath} as a feature extractor. 
Provided with $\mathcal{X}^r$, we utilize a GMM to learn the distribution of the environment surrounding rare-type nuclei, modeled as $p(\mathcal{P}^r) = p(\mathcal{X}^r) = \mathcal{N}(\mu^r, \Sigma^r)$ where $\mu^r$ and $\Sigma^r$ denote the mean and covariance matrix of the rare-type feature vectors $\mathcal{X}^r$, respectively. This model is then used to predict the likelihood of the major-type and background feature vectors by measuring how closely they resemble rare-type feature vectors. These likelihood are expressed as $p(\mathcal{P}^m ~| ~\mathcal{P}^r) = \mathcal{N}(\mathcal{X}^m ~| ~\mu^r, \Sigma^r)$ and $p(\mathcal{P}^b ~| ~\mathcal{P}^r) = \mathcal{N}(\mathcal{X}^b ~| ~\mu^r, \Sigma^r)$, respectively. In other words, we conjecture that the neighboring environment of $\alpha_i^m$ or $\alpha_i^b$ is similar to that of rare-type nuclei if $p({\phi}_i^m ~| ~\mathcal{P}^r)$ or $p({\phi}_i^b ~| ~\mathcal{P}^r)$ is high.

Next, we sample candidate locations to insert rare-type nuclei from major-type and background patches based on the joint likelihood $\hat{\phi}^t \sim p(\mathcal{P}^m, \mathcal{P}^b ~| ~\mathcal{P}^r)$, where $t$ is either $m$ or $b$. Here, $\hat{\phi}^t$ are the patches that share similar neighboring environments with the rare-type patches $\mathcal{P}^r$.
For each $\hat{\phi}^t$, we choose a specific rare-type nucleus that best matches the new neighboring environments based upon the probability distributions given by: 

\begin{equation}
    p({\phi_i^r | \hat{\phi}^t}) = \frac{e^{-w_i \times D(\hat{\phi}^t,\phi_i^r)}}{\sum_{k=1}^{N_r} e^{-w_k \times D(\hat{\phi}^t,\phi_k^r)}}
\end{equation}

\noindent where $D(\hat{\phi}^t, \phi_i^r) = ||\hat{x}^t - x_i^r||_2$, $\hat{x}^t$ is the feature vectors of $\hat{\phi}^t$, and $||.||_2$ denotes the L2 distance. $w_i$ is the selection count of $\alpha_i^r$, dynamically updated during the Copy-and-Replace and Copy-and-Paste operation to discourage repeated use of the same rare-type nucleus.
Sampling the most likely rare-type patches for the major-type and background locations, designated as $\phi^{r|m}$ and $\phi^{r|b}$, we insert the corresponding rare-type nuclei into these selected locations, respectively.

\begin{figure*}[htbp]
    \centering
    \includegraphics[width=\linewidth]{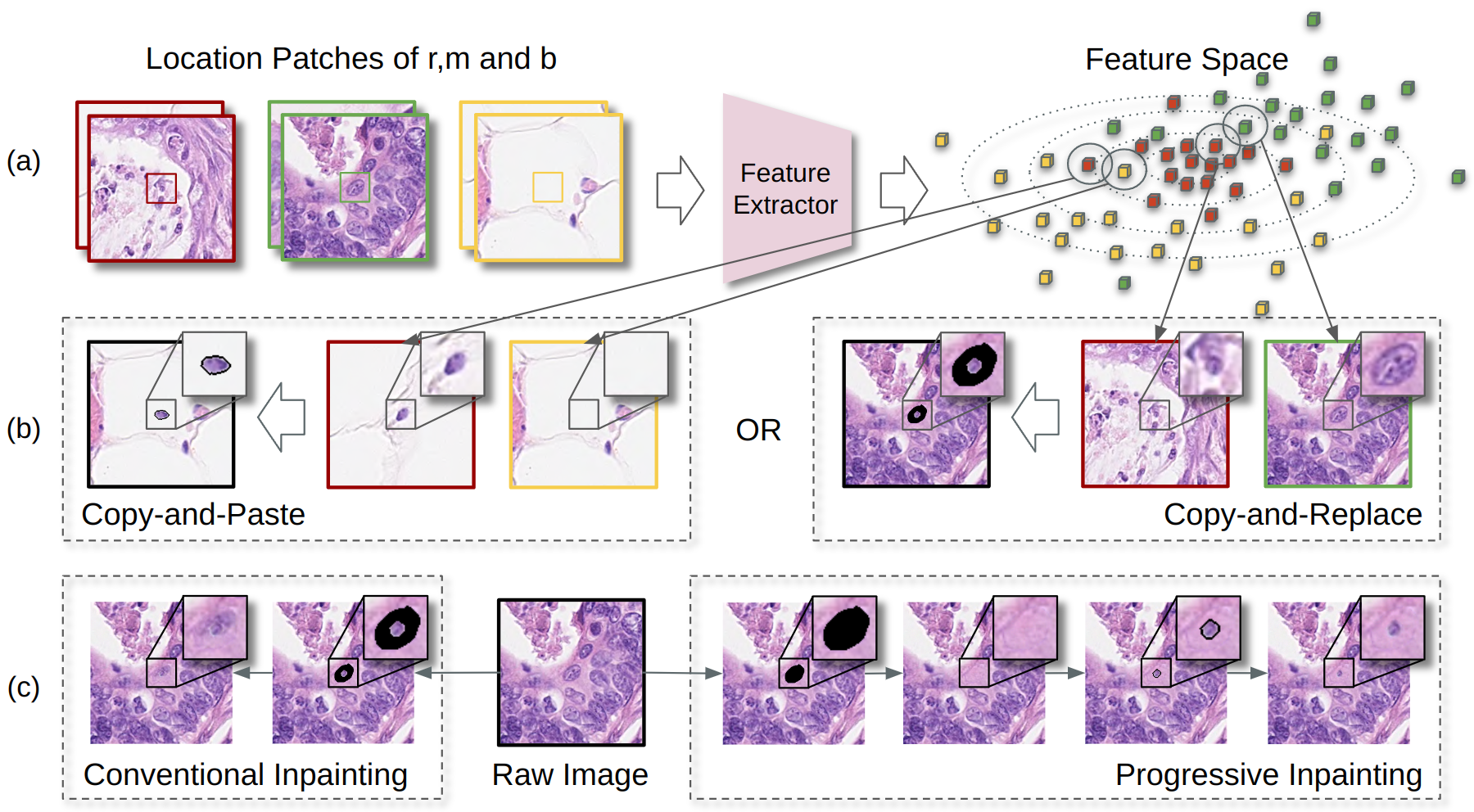} 
    \caption{
    (a) Procedure for forming a feature space for local patches. Rare-type (r in red), major-type (m in green), and background (b in yellow) locations are obtained and fed into a pre-trained feature extractor to extract the corresponding feature vectors (rare-type feature as red cube, major-type as green cube, and background feature as yellow cube), forming a feature space. 
    (b) Procedure of context-aware probabilistic sampling for selecting candidate locations. A major-type (green cube) or background (yellow cube) feature vector is sampled, which is close to the rare-type feature distribution (red cubes). An instance of rare-type nuclei, closest to the target locations in the feature space, is sampled and is used to replace the major-type nucleus (Copy-and-Replace) or paste into the background location (Copy-and-Paste).
    (c) Procedure of the progressive inpainting strategy. Selecting a major-type location, the progressive inpainting strategy removes the major-type nucleus, inpaints the blank region using a diffusion model, places a rare-type nucleus with an immediate one-layer contour from the original environment and two-layers of a transition zone, and inpaints the transition zone using the diffusion model. In contrast, the conventional method places the rare-type nucleus in the blank region and applies inpainting directly to the blank region.
    }
    \label{fig:procedure}
\end{figure*}

\begin{figure*}[htbp]
    \centering
    \includegraphics[width=\linewidth]{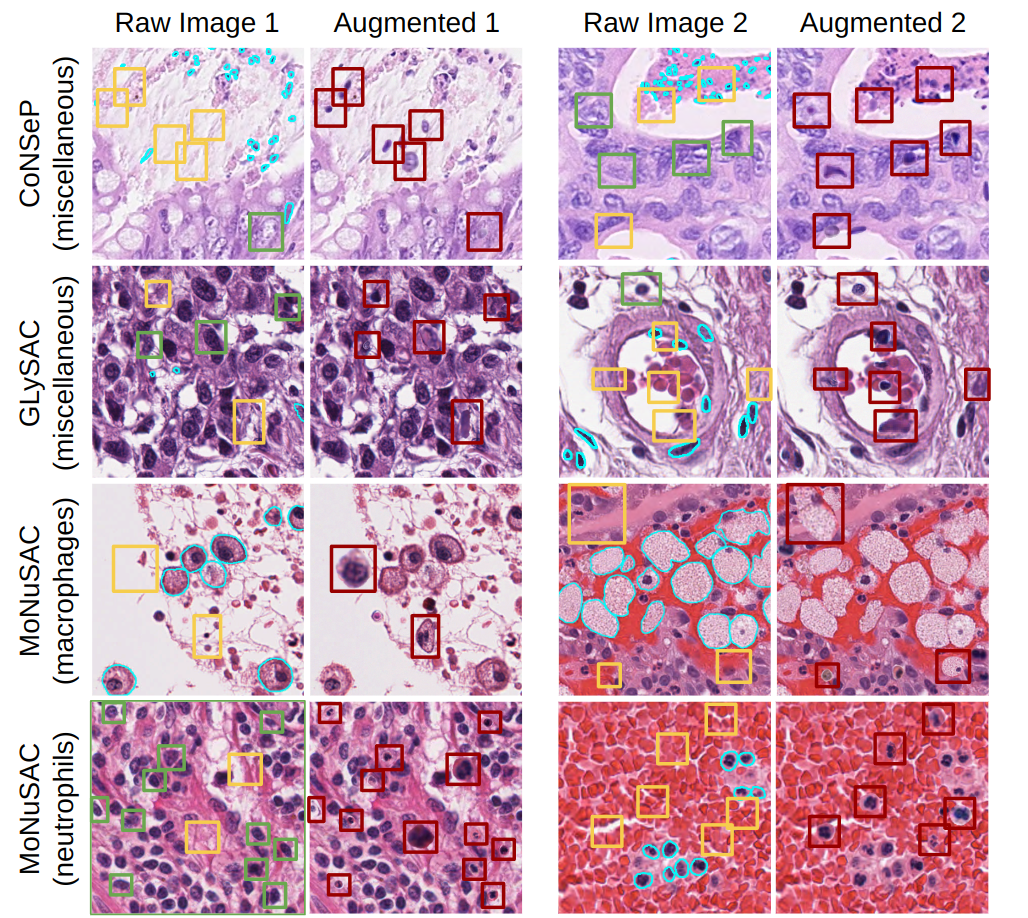} 
    \caption{
    Visual assessment of the new rare-type nuclei appearance. In the raw images, original rare-type nuclei are contoured in sky blue, major-type positions are green boxes, and background-type positions are yellow boxes. In the augmented images, new rare-type nuclei are highlighted by red boxes.
    }
    \label{fig:ope_examples}
\end{figure*}

For the major-type location, we remove the existing major-type nucleus and replace it with a new rare-type nucleus. This process can be understood as a Copy-and-Replace operation. As for the background location, we directly insert a new rare-type nucleus, which can be described as a Copy-and-Paste operation. Once the operation is complete, the major-type nucleus or background location will no longer be considered. The entire procedure is illustrated in Fig.~\ref{fig:procedure}. In addition, we present two local views for each dataset to show the regions before and after the augmentation operations in Fig.~\ref{fig:ope_examples}. Specifically, in the case of macrophages in the MoNuSAC dataset, we only apply a Copy-and-Paste operation due to their large average size.

\subsection{Progressive Inpainting with Diffusion Model}
Although the above approach effectively identifies the patches with similar neighboring environments, the simple placement of rare-type nuclei can still result in artifacts or discontinuities surrounding the inserted nuclei. Generative models such as GAN~\cite{gan} or diffusion models~\cite{dm, ddpm} can be adopted to fill in the artificial, discontinuous, or missing regions. However, we have observed that such direct inpainting can sometimes lead to blurry or obscure boundaries around the inserted nuclei. As shown in Fig.\ref{fig:procedure}b, existing diffusion models tend to harmonize the inpainted regions with their surroundings, which can result in a loss of sharpness at the nuclear boundaries. To address such issues, we introduce a progressive inpainting strategy that can preserve the integrity of the inserted nuclei while seamlessly integrating them into the surrounding environments. The proposed method conducts inpainting in two stages with distinct constraints applied to the diffusion model. 
In the first stage, for a major-type location $\hat{\phi}^m$, we erase the region previously occupied by a major-type nucleus $\hat{\alpha}^m$, creating a blank space, and use a diffusion model to inpaint this blank region. For a background location $\hat{\phi}^b$, we skip the first stage and proceed to the second stage.
In the second stage, we position rare-type nuclei, $\alpha^{r|m}$ and $\alpha^{r|b}$, at the center of the major-type and background patches, $\hat{\phi}^m$ and $\hat{\phi}^b$, respectively. To preserve the original neighborhood information, the immediate one-layer contour surrounding $\alpha^{r|m}$ and $\alpha^{r|b}$ is placed together.
We then create a transition zone around the inserted rare-type nuclei by removing the second and third-layer contours surrounding them. The diffusion model is utilized once again to inpaint the transition zone, achieving a seamless transition with the surrounding environments.

\subsection{Diffusion Model: Solving Inverse Problems}
Diffusion models define the forward diffusion process as a Markov Chain where the data is gradually corrupted by adding noise, while their generative process reverses this noising process to recover the original data. Let latents ${\mathbf{x}}_t$, $\forall t \in [0, T]$ be a continuous diffusion process such that ${\mathbf{x}}_0 \sim p_{data}$ is data distribution and ${\mathbf{x}}_T \sim \mathcal{N}(0,\mathbf{I})$ is isotropic Gaussian distribution.~\cite{ddpm} defines the forward process as:
\begin{equation}
    q(\boldsymbol{x}_t|\boldsymbol{x}_{t-1}) := \mathcal{N}(\boldsymbol{x}_{t-1}; \sqrt{1-\beta_t} \boldsymbol{x}_{t-1}, \beta_t \mathbf{I}).
\label{eq:forward}
\end{equation}
\noindent The tractable posterior of~\eqref{eq:forward} is:
\begin{equation}
    q(\boldsymbol{x}_{t-1}|\boldsymbol{x}_{t}, \boldsymbol{x}_0) := \mathcal{N}(\boldsymbol{x}_{t-1}; \tilde{\boldsymbol{\mu}_t}(\mathbf{x}_t(\mathbf{x_0, \mathbf{\epsilon}}), t), \tilde{\beta_t} \mathbf{I}).
\label{eq:forward_posterior}
\end{equation}
\noindent $\epsilon \sim \mathcal{N}(0,\mathbf{I})$, ${\beta}_t$ and $\tilde{\beta}_t$ are predefined hyperparameters. The model progressively removes noise in a reverse process:
\begin{equation}
    p_{\theta}(\mathbf{x}_{t-1} | \mathbf{x}_t) = \mathcal{N} \left( \mathbf{x}_{t-1} \middle| \mu_{\theta} (\mathbf{x}_t, \epsilon_{\theta}(\mathbf{x_t}, t), \tilde{\beta}_t \mathbf{I} \right)
\end{equation}
\noindent where $\mathbf{\epsilon_{\theta}}$ is an approximator to predict $\epsilon$ from $\mathbf{x_t}$. The model is trained by decreasing the KL divergence between two Gaussians, and a simple objective function is given by:
\begin{equation}
\mathcal{L} = E_{\mathbf{x_t},t,\epsilon \sim \mathcal{N}(0,1)} \left[ \left\| \epsilon - \epsilon_\theta(\mathbf{x_t}, t) \right\|_2^2 \right].
\end{equation}
\noindent With a pre-trained approximator $\epsilon_\theta$, we can sample a image from $\mathbf{x}_T$ that match the measurement $\mathbf{y}$ during $t\sim[T, 1]$, according to~\cite{dps}:
\begin{equation}
\hat{\mathbf{x}}_0 = \left( \mathbf{x}_t - \sqrt{1 - \bar{\alpha}_t} \, \boldsymbol{\epsilon}_{\theta}(\mathbf{x}_t) \right) / \sqrt{\bar{\alpha}_t}
\label{eq:tweedie}
\end{equation}
\begin{equation}
\boldsymbol{x}_{t-1}' = \sqrt{\bar{\alpha}_{t-1}} \hat{\mathbf{x}}_0 + \sqrt{1 - \bar{\alpha}_{t-1}} \, \boldsymbol{\epsilon}_{\theta}(\boldsymbol{x}_t)
\end{equation}
\begin{equation}
\boldsymbol{x}_{t-1} = \boldsymbol{x}_{t-1}' - \gamma \nabla_{\mathbf{x}_t} \left\| \mathbf{y} - \mathcal{A}(\hat{\mathbf{x}}_0) \right\|_2^2
\end{equation}
\noindent $\hat{\mathbf{x}}_0$ in \eqref{eq:tweedie} is the posterior mean by Tweedie’s formula~\cite{twe}. In the case of inpainting~\cite{repaint, mcg}, the measurement $\mathbf{y}$ is the masked condition image, and $\mathcal{A}(\cdot)$ denotes the masking operation.

\begin{algorithm}
\caption{NucleiMix Augmentation}
\begin{algorithmic}[1]
\Require Pathology image dataset $\mathcal{D}$ with a number of rare-type nuclei $\mathcal{A}^{r} = \{\alpha_i^r | i=1,...,N_r\}$, major-type nuclei $\mathcal{A}^{m} = \{\alpha_i^m | i=1,...,N_m\}$, and background locations $\mathcal{A}^{b} = \{\alpha_i^b | i=1,...,N_b\}$.

\State $\mathcal{P}^{r} = \{\phi_i^r | i=1, ..., N_r\} \leftarrow CropPatch(\mathcal{A}^{r})$
\State $\mathcal{P}^{m} = \{\phi_i^m | i=1, ..., N_m\} \leftarrow CropPatch(\mathcal{A}^{m})$
\State $\mathcal{P}^{b} = \{\phi_i^b | i=1, ..., N_b\} \leftarrow CropPatch(\mathcal{A}^{b})$

\LeftComment{Crop patches using $\mathcal{A}^{r}$, $\mathcal{A}^{m}$, and $\mathcal{A}^{b}$.}

\State $\mathcal{X}^r=\{x_i^r\in \mathbb{R}^p, i=1, ..., N_r\} \leftarrow ExtractFeatures(\mathcal{P}^{r})$
\State $\mathcal{X}^m=\{x_i^m\in \mathbb{R}^p, i=1, ..., N_m\} \leftarrow ExtractFeatures(\mathcal{P}^{m})$
\State $\mathcal{X}^b=\{x_i^b\in \mathbb{R}^p, i=1, ..., N_b\} \leftarrow ExtractFeatures(\mathcal{P}^{b})$

\LeftComment{Extract feature vectors from $\mathcal{P}^{r}$, $\mathcal{P}^{m}$, and $\mathcal{P}^{b}$.}

\State Use GMM to model the distribution of $p(\mathcal{P}^r)$ = $p(\mathcal{X}^r) = \mathcal{N}(\mu^r, \Sigma^r)$
\State Predict the conditional joint likelihood of major-type patches and background-type patches $p(\mathcal{P}^m, \mathcal{P}^b ~| ~\mathcal{P}^r) = \mathcal{N}(\mathcal{X}^m, \mathcal{X}^b ~| ~\mu^r, \Sigma^r)$
\For{$j=1$ to $k$}
    \State Sample $\hat{\phi}^t_j \sim p(\mathcal{P}^m, \mathcal{P}^b ~| ~\mathcal{P}^r)$, where $t$ is either $m$ or $b$
    \State Sample the most likely rare-type patch $\phi^{r|t} \sim p({\phi_i^r | \hat{\phi}_j^t}) = \frac{e^{-w_i \times D(\hat{\phi}_j^t,\phi_i^r)}}{\sum_{k=1}^{N_r} e^{-w_k \times D(\hat{\phi}_j^t,\alpha_k^r)}}$
    
    \If{$\hat{\phi}^t_j$ is major-type patch ($t$ = $m$)}
        \State Copy-and-Replace: copy the $\alpha^r$ at the center of $\phi^{r|m}$ to replace 
 the $\alpha^m$ at the center of $\phi^m_j$
    \ElsIf{$\hat{\phi}^t_j$ is background-type patch ($t$ = $b$)}
        \State Copy-and-Paste: copy the $\alpha^r$ at the center of $\phi^{r|m}$ to paste at the center of $\phi^m_j$
    \EndIf
\EndFor
\end{algorithmic}
\label{alg}
\end{algorithm}

\section{Experiments}
\subsection{Datasets}
\subsubsection{Dataset for Diffusion Model Training}
A diffusion model is trained on a collection of pathology image patches of 12 different organs from TCGA (https://github.com/aleju/imgaug), including 1) 5k bladder urothelial carcinoma; 2) 5k breast invasive carcinoma; 3) 5k cervical squamous cell carcinoma and endocervical adenocarcinoma; 4) 5k colon adenocarcinoma; 5) 2460 kidney chromophobe; 6) 5k liver hepatocellular carcinoma; 7) 5k lung adenocarcinoma; 8) 4090 pancreatic adenocarcinoma; 9) 5k prostate adenocarcinoma; 10) 1880 rectum adenocarcinoma; 11) 5k stomach adenocarcinoma; 12) 1640 uveal melanoma. Each image patch has a spatial size of 256$\times$256 pixels and a resolution of 0.5{{\textmu}m}$\times$0.5{{\textmu}m} per pixel. 

\subsubsection{Dataset for Nuclei Instance Segmentation}
We employ three public datasets for nuclei instance segmentation such as CoNSeP, GLySAC, and MoNUSAC. Table ~\ref{tab:tab1} depicts the details of these datasets. 
These datasets include at least one rare-type. A nucleus type is defined as rare-type if meets at least one of the following conditions: 1) Frequency condition: the number of nuclei belonging to a nucleus type is less than 5\% of the total number of nuclei in the dataset and 2) Miscellaneous condition: the nucleus type is explicitly labeled as "Miscellaneous". Specifically, the miscellaneous class appears in CoNSeP and GLySAC. Both contain multiple nucleus class labels and/or unidentified or ambiguous class labels. Hence, the precise composition of each nucleus type within the miscellaneous category is unknown, and their individual proportions are likely smaller than other clearly defined classes. 

CoNSeP, which stands for Colorectal Nuclear Segmentation and Phenotypes dataset, was extracted from 16 colorectal adenocarcinoma whole slide images (WSIs) and contains 41 image patches of size 1000$\times$1000 pixels, digitized at 40$\times$ magnification. The dataset has four classes: epithelial, inflammatory, spindle, and miscellaneous. Among these four classes, miscellaneous is considered as a rare-type, which includes necrotic cells, mitotic cells, and cells that cannot be categorized.
The dataset is partitioned into a training set of 27 images and a testing set of 14 images. 

GLySAC, known as Gastric Lymphocyte Segmentation And Classification, was derived from 8 gastric adenocarcinoma WSIs that were digitized at 40$\times$ magnification. This dataset contains 59 image patches, of which 34 patches are designated as the training dataset and the remaining 25 patches are the test dataset. All nuclei in the dataset are classified into three types: lymphocytes, epithelial, and miscellaneous. Lymphocytes and epithelial nuclei are major-type nuclei, while miscellaneous is a rare-type. Miscellaneous contains nuclei other than lymphocytes and epithelial nuclei such as stromal nuclei and endothelial nuclei, and other uncategorized nuclei.

MoNuSAC (Multi-organ Nuclei Segmentation And Classification) comprises 209 annotated image patches of size ranging from 81$\times$113 to 1422$\times$2162 pixels. The image patches are split into a training set of 168 patches and a test set of 41 patches. These patches are sourced from various organs, including the breast, kidney, lung, and prostate. The annotated nuclei are grouped into four types: epithelial, lymphocytes, macrophages, and neutrophils. Major-type nuclei include epithelial and lymphocytes. Rare-type nuclei consist of macrophages and neutrophils.

\subsection{Experimental Design}
To investigate the effectiveness of NucleiMix, we conduct nuclei instance segmentation with and without data augmentation using NucleiMix and compare the performance between them on the three public datasets (CoNSeP, GLySAC, and MoNuSAC). For each of these datasets, we augment nuclei data by inserting k rare-type nuclei, either in place of major-type nuclei or at the background locations. 
We set k = 200, 400, 600, and 800 for all three datasets. For nuclei instance segmentation, we employ two popular backbone models SONNET~\cite{glysac} and PointNu~\cite{pointnu}. Both models are separately trained and evaluated on the augmented datasets with varying k to assess the impact of the augmented instances on the performance. Moreover, we compare NucleiMix with three other data augmentation methods such as CutMix~\cite{cutmix}, CopyPaste~\cite{copypaste}, and GradMix~\cite{gradmix}. Each of these methods is used in the same manner as NucleiMix to assess the performance of nuclei instance segmentation on the three public datasets.
To choose the optimal value of k, we conduct experiments with the training sets augmented by NucleiMix. Based upon the overall segmentation and classification performance, k is set to 600 and fixed for all comparison experiments with other augmentation methods.

\subsection{Implementation and Training Details}
We used 2 RTX6000 GPUs to train a diffusion model with 300k steps using 50,070 image patches of 256$\times$256 pixels, loaded with the 256$\times$256 diffusion pre-trained weight. For image sampling, we applied 250 DDIM steps~\cite{ddim} and the Manifold Constraint Gradient(MCG)~\citep{mcg} correction term to boost the inpainting performance.

We trained two nuclei segmentation and classification models, SONNET and PointNu, on the original datasets (without data augmentation) for 100 epochs, forming baseline models. Starting from the 50th checkpoint of these baseline models, we trained the two models on the augmented datasets for an additional 50 epochs. All models were trained with Adam optimizer (${\beta}_1 = 0.9, {\beta}_2 = 0.999, {\epsilon} = 1.0e^{-8}$) and cosine annealing warm restarts scheduler with $eta\_min=1.0e^{-5}$ and $T_0 = 25$. The learning rate was initially set to $1.0e^{-4}$, decreasing to $1.0e^{-5}$ after every 25 epochs and restarting from $1.0e^{-4}$. The image patches were resized to 512$\times$512 pixels.

For SONNET, we used 2 RTX6000 GPUs to train a baseline model, with a batch size for each GPU set to 16 for the first 50 epochs and 4 for the second 50 epochs. To train it on the augmented datasets, a batch size for each GPU was set to 4 for 50 epochs. As for PointNu, we used 1 RTX6000 GPU with a batch size of 16. Traditional on-the-fly data augmentation techniques, including random horizontal and vertical flips, Gaussian blurring, Gaussian noise, color change in hue, saturation, and modification of contrast, were utilized during training. These were implemented using Aleju library (https://github.com/aleju/imgaug). To determine the checkpoint for test evaluation, we selected the best-performing one among the last five training epochs for both models.

\subsection{Evaluation metrics}

To evaluate the performance of nuclei segmentation, we employ five evaluation metrics: 1) DICE score to measure the overlap between the predicted segmentation and the ground truth~\cite{dice}, 2) Aggregated Jaccard Index (AJI) to evaluate instance segmentation by computing the intersection-over-union (IoU) for matched nuclei and penalizing both false positives and false negatives~\cite{aji}, 3) Detection Quality (DQ) matches predicted instances to ground truth using a IoU threshold ($>$0.5), and quantifies the harmonic mean of precision and recall at the object level, to show how accurately individual nuclei are detected, 4) Segmentation Quality (SQ) measures how well the segmentation matches the ground truth once the objects are correctly detected, 5) Panoptic Quality (PQ) is a combination of DQ and SQ, robust measurement of how well the objects are detected and how well they are segmented~\cite{dq_sq_pq}, 6) Boundary Intersection-over-Union (b-IoU) to measure the boundary of the accuracy, we set boundary pixel width as 0.5\% of an image diagonal for the dataset, as the paper suggested~\cite{b_iou}. For nuclei classification, we utilized 1) Detection Quality $F_d$ score to measure how well the objects are detected and classified~\cite{dq_sq_pq}, 2) Accuracy ($Acc$) to measure the proportion of correct predictions, and 3) $F_1$ score for each nuclei type to measure how well the model identifies positive instances~\cite{f1}, denote as $F^t_c$.

\begin{table*}[htbp]
\caption{Data Distribution} 
\label{tab:tab1}
\begin{center}    
\resizebox{\linewidth}{!}{
\begin{tabular}{ccclllllll} 
\hline
\multirow{2}{*}{Dataset} & \multirow{2}{*}{Type} &  \multirow{2}{*}{Baseline}    &   \multicolumn{4}{c}{NucleiMix}  & CutMix &  CopyPaste   &   GradMix
\\ 
 &  & &   k=200   &   k=400   &   k=600   &   k=800   & k=600 &   k=600   &   k=600
\\
\hline
\multirow{4}{7em}{CoNSeP\\(Miscellaneous)} & miscellaneous & 371 & 569 (+198) & 769 (+398) & 967 (+596) & 1163 (+792) & 968 (+597) & 970 (+599) & 971 (+600) \\
       & inflammatory & 3941 & 3938 (-3) & 3935 (-6) & 3931 (-10) & 3931 (-10) & 3940 (-1) & 3941 (-0) & 3772 (-169) \\
       & epithelial  & 5537 & 5457 (-80) & 5400 (-137) & 5352 (-185) & 5273 (-264) & 5537 (-0) & 5537 (-0) & 5306 (-231) \\
       & spindle     & 5706 & 5690 (-16) & 5672 (-34) & 5659 (-47) & 5643 (-63) & 5704 (-2) & 5706 (-0) & 5506 (-200) \\
\hline
\multirow{3}{7em}{GLySAC\\(Miscellaneous)} & miscellaneous & 3386 & 3586 (+200) & 3786 (+400) & 3986 (+600) & 4183 (+797) & 3982 (+596) & 3985 (+599) & 3986 (+600) \\
       & inflammation & 7409 & 7356 (-53) & 7300 (-109) & 7254 (-155) & 7214 (-195) & 7405 (-4) & 7409 (-0) & 6998 (-411) \\
       & epithelial & 7154 & 7091 (-63) & 7033 (-121) & 6987 (-167) & 6953 (-201) & 7152 (-2) & 7152 (-2) & 6965 (-189) \\
\hline
\multirow{4}{7em}{MoNuSAC\\(Macrophages)} & epithelial & 12121 & 12118 (-3) & 12113 (-8) & 12105 (-16) & 12096 (-25) & 12040 (-81) & 12077 (-44) & 11736 (-385) \\
       & lymphocyte & 12524 & 12512 (-12) & 12506 (-18) & 12496 (-28) & 12493 (-31) & 12246 (-278) & 12337 (-187) & 11891 (-633) \\
       & macrophages & 422 & 620 (+198) & 818 (+396) & 1018 (+596) & 1217 (+795) & 1013 (+591) & 1015 (+593) & 1021 (+599) \\
       & neutrophils & 541 & 540 (-1) & 540 (-1) & 539 (-2) & 539 (-2) & 522 (-19) & 518 (-23) & 540 (-1) \\
\hline
\multirow{4}{7em}{MoNuSAC\\(Neutrophils)} & epithelial & 12121 & 12091 (-30) & 12053 (-68) & 12026 (-95) & 11997 (-124) & 12120 (-1) & 12077 (-44) & 11672 (-449) \\
       & lymphocyte & 12524 & 12478 (-46) & 12445 (-79) & 12414 (-110) & 12393 (-131) & 12511 (-13) & 12337 (-187) & 12373 (-151) \\
       & macrophages & 422 & 422 (-0) & 422 (-0) & 422 (-0) & 422 (-0) & 422 (-0) & 422 (-0) & 422 (-0) \\
       & neutrophils & 541 & 741 (+200) & 941 (+400) & 1141 (+600) & 1340 (+799) & 1140 (+599) & 1140 (+599) & 1141 (+600) \\
\hline
\multirow{4}{7em}{MoNuSAC\\(Macrophages and Neutrophils)} 
       & epithelial & 12121 & 12085 (-36) & 12031 (-90) & 11951 (-170) & 11907 (-214) & 12036 (-85) & 12077 (-44) & 11347 (-774) \\
       & lymphocyte & 12524 & 12464 (-60) & 12384 (-140) & 12350 (-174) & 12304 (-220) & 12187 (-337) & 12337 (-187) & 11680 (-844) \\
       & macrophages & 422 & 623 (+200) & 822 (+400) & 1017 (+595) & 1215 (+793) & 1015 (+593) & 1015 (+593) & 1021 (+599) \\
       & neutrophils & 541 & 737 (+196) & 936 (+395) & 1138 (+597) & 1339 (+798) & 1071 (+530) & 1086 (+545) & 1140 (+599) \\
\hline
\end{tabular}
}
\end{center}
\end{table*}

\begin{table*}[htbp]
\caption{Distribution of Newly Added Rare-type Nuclei in Major-type and Background Positions} 
\label{tab:tab2}
\begin{center}    
\resizebox{0.7\linewidth}{!}{
\begin{tabular}{lcccccc} 
\hline
 Dataset & Type &  Total & k=200   &   k=400   &   k=600   &   k=800    
\\
\hline
\multirow{2}{12em}{CoNSeP (Miscellaneous)} 
       & $\alpha^m$ & 15184 & 99 & 177 & 242 & 337   \\
       & $\alpha^b$ & 12031 & 101 & 223 & 358 & 463  \\
\hline
\multirow{2}{12em}{GLySAC (Miscellaneous)} 
       & $\alpha^m$ & 14563 & 116 & 230 & 322 & 396  \\
       & $\alpha^b$ & 15440 & 84 & 170 & 278 & 404 \\
\hline
\multirow{2}{12em}{MoNuSAC (Macrophages)}
       & $\alpha^m$ & 0 & - & - & - & - \\
       & $\alpha^b$ & 36750 & 200 & 400 & 600 & 800 \\
\hline
\multirow{2}{12em}{MoNuSAC (Neutrophils)} 
       & $\alpha^m$ & 24645 & 76 & 147 & 205 & 255 \\
       & $\alpha^b$ & 36750 & 124 & 253 & 395 & 545 \\
\hline
\multirow{2}{12em}{MoNuSAC (Macrophages and Neutrophils)} 
       & $\alpha^m$ & 24645 & 96 & 230 & 344 & 442 \\
       & $\alpha^b$ & 36750 & 304 & 570 & 856 & 1178 \\
\hline
\end{tabular}
}
\end{center}
\end{table*}

\section{Results}
We explored the impact of augmenting rare-type nuclei on three datasets: CoNSeP~\cite{consep}, GLySAC~\cite{glysac}, and MoNuSAC~\cite{monusac}. We use k to denote the number of synthesized nuclei added to the original dataset. In CoNSeP and GLySAC, there is only one type of rare-type nuclei, which is miscellaneous. k number of new miscellaneous nuclei were generated and added to each of the datasets. In the case of MoNuSAC, there are two types of rare-type nuclei: neutrophils and macrophages. We generated three kinds of augmented datasets: 1) only adding k synthesized neutrophils; 2) only adding k synthesized macrophages; 3) adding k synthesized macrophages and k synthesized neutrophils together. 
Details of the nuclei label distribution in the original and augmented datasets are available in Table \ref{tab:tab1}, and Table~\ref{tab:tab2} shows the distribution of newly added rare-type nuclei in the major-type and background-type positions.
We compared NucleiMix with three other methods: CutMix~\cite{cutmix}, CopyPaste~\cite{copypaste}, and GradMix~\cite{gradmix} in k = 600, and evaluated the nuclei segmentation and classification performance with and without the augmented datasets on two popular models: SONNET~\cite{glysac} and PointNu~\cite{pointnu}. 

\begin{figure*}[htbp]
    \centering
    \includegraphics[width=1.0\linewidth]{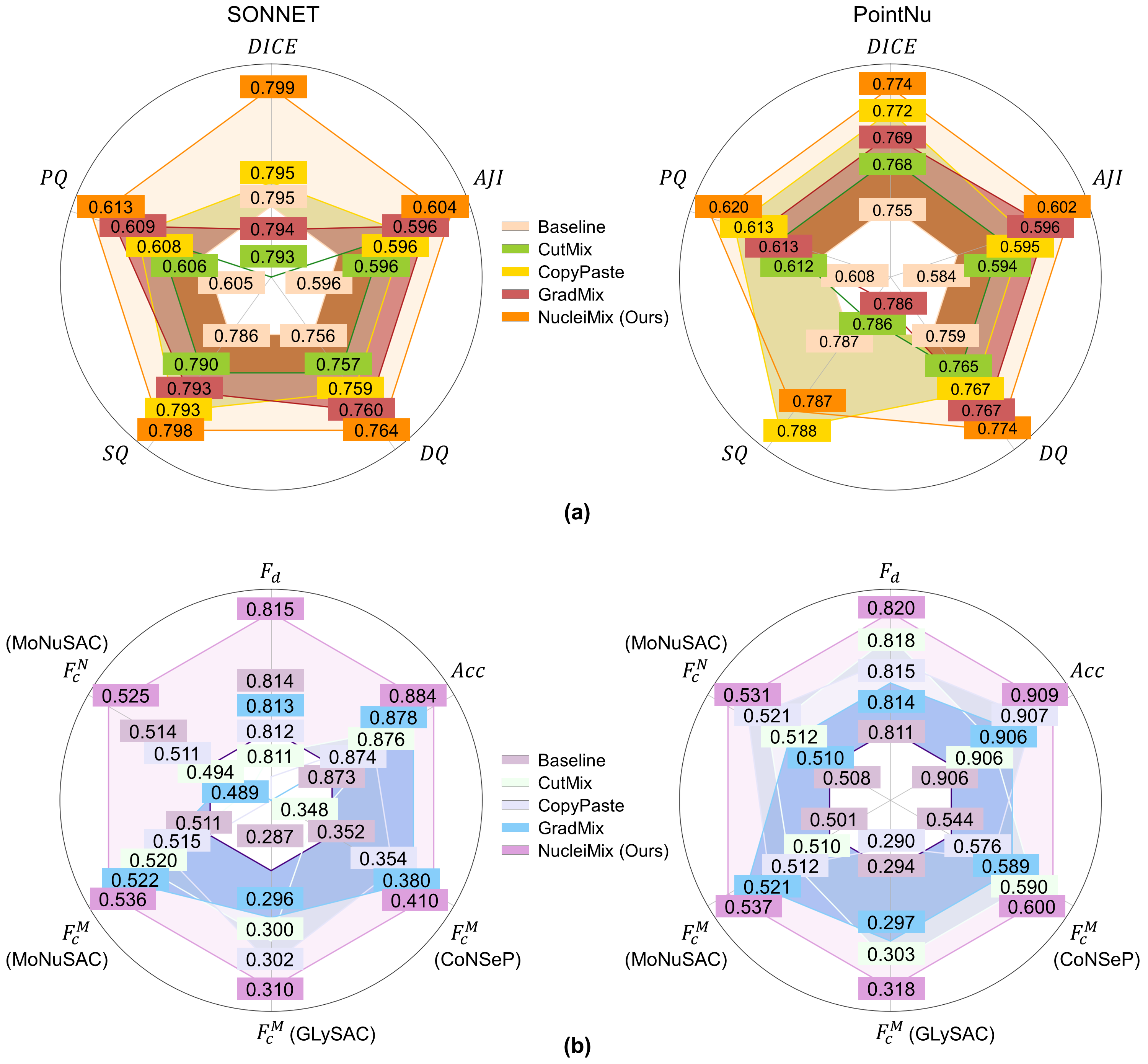} 
    \caption{Radar charts for (a) nuclei segmentation and (b) nuclei classification using SONNET and PointNu, respectively, among baseline, CutMix, CopyPaste, Gradmix, and NucleiMix. The numbers represent the average scores across three datasets (CoNSeP, GLySAC, and MoNuSAC). }
    \label{fig:radar}
\end{figure*}

Fig. \ref{fig:radar}a demonstrates the average performance of nuclei segmentation for SONNET and PointNu using NucleiMix, the baseline (without augmentation), and other augmentation methods (CutMix, CopyPaste, and GradMix) across the three datasets (CoNSeP, GLySAC, and MoNuSAC). 
In comparison to the baseline and the three augmentation methods, the strength of NucleiMix was prominent. For both SONNET and PointNu, the adoption of NucleMix, on average, achieved the best performance in all evaluation metrics except SQ for PointNu. 
For example, using NucleiMix, SONNET obtained 0.799 DICE, 0.604 AJI, 0.764 DQ, 0.798 SQ, and 0.613 PQ, while PointNu attained 0.774 DICE, 0.602 AJI, 0.774 DQ, 0.787 SQ, and 0.620 PQ. 
Among the three other augmentation methods, CopyPaste and GradMix were comparable to each other, achieving the second or third-best performance in most of the evaluation metrics. 
PointNu with CopyPaste, in particular, achieved the highest SQ of 0.788 but obtained the second-best DICE and PQ and third-best AJI and DQ. 
CutMix was shown to be the poorest among the augmentation methods. 
Without data augmentation (baseline), both SONNET and PointNu attained the worst performance in all evaluation metrics except DICE for SONNET and SQ for PointNu.

Moreover, we measured and averaged the results of nuclei classification for SONNET and PointNu with and without augmentation methods across the three datasets (Fig. \ref{fig:radar}b). 
NucleiMix outperformed other augmentation methods as well as the baseline for both SONNET and PointNu. With SONNET, it achieved 0.815 $F_d$, 0.884 Acc, 0.410 $F^M_c$ in CoNSeP, 0.310 $F^M_c$ in GLySAC, 0.536 $F^M_c$ in MoNuSAC, and 0.525 $F^N_c$ in MoNuSAC. With PointNu, it obtained 0.820 $F_d$, 0.909 Acc, 0.600 $F^M_c$ in CoNSeP, 0.318 $F^M_c$ in GLySAC, 0.537 $F^M_c$ in MoNuSAC, and 0.531 $F^N_c$ in MoNuSAC. 
The performance of the other three augmentation methods varied depending on the evaluation metrics, and thus there was no clear second-best-performing model. 
The baseline was, on the other hand, generally inferior to NucleiMix and other augmentation methods.

\begin{figure*}[h]
    \centering
    \includegraphics[width=1.0\linewidth]{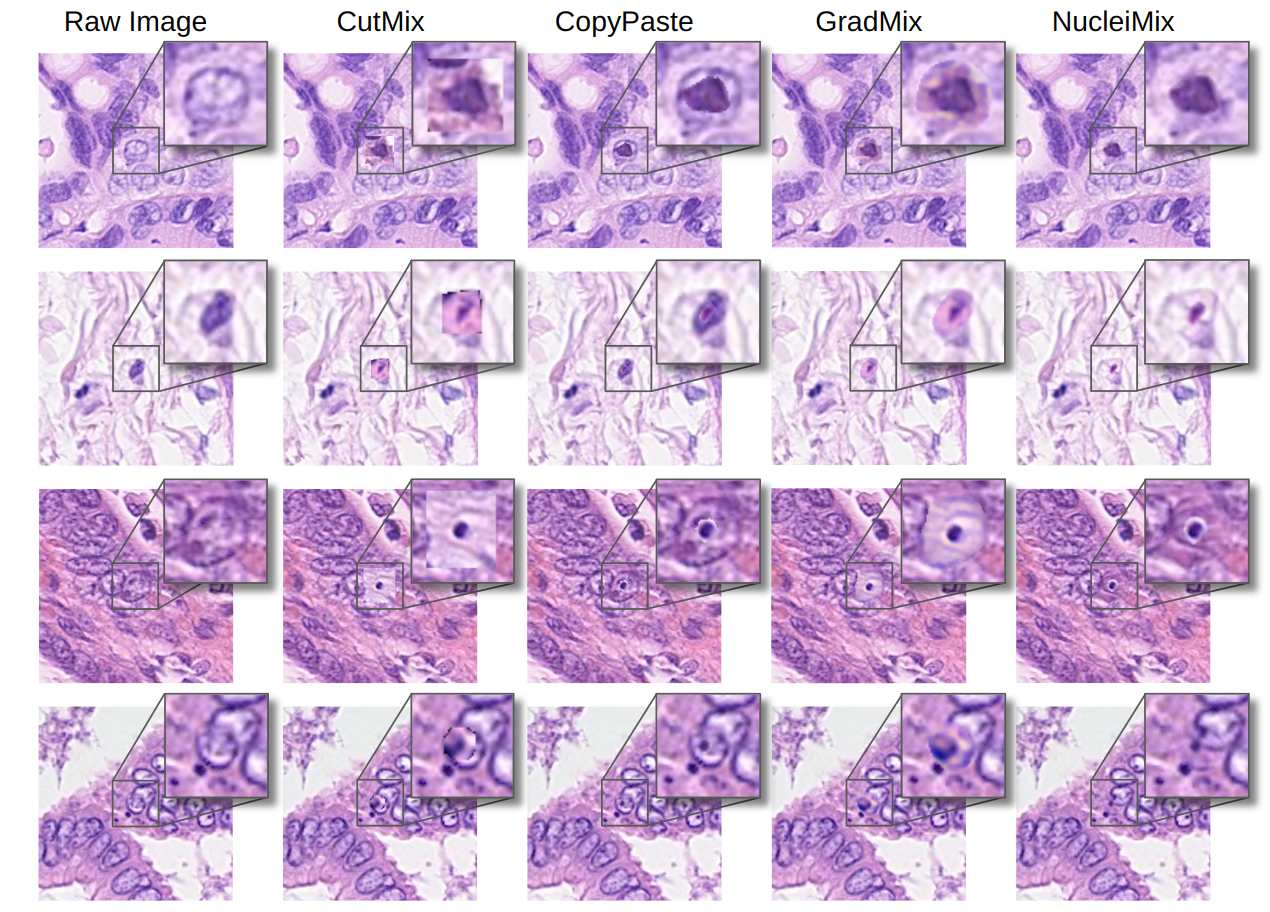}
    \caption{Visual comparison of nuclei augmentation between NucleiMix with other augmentation methods. The first two rows show isolated nuclei and the latter two rows present touched nuclei.}
    \label{fig:comparisons}
\end{figure*}

\begin{figure*}[!]
    \centering
    \includegraphics[width=0.99\linewidth] {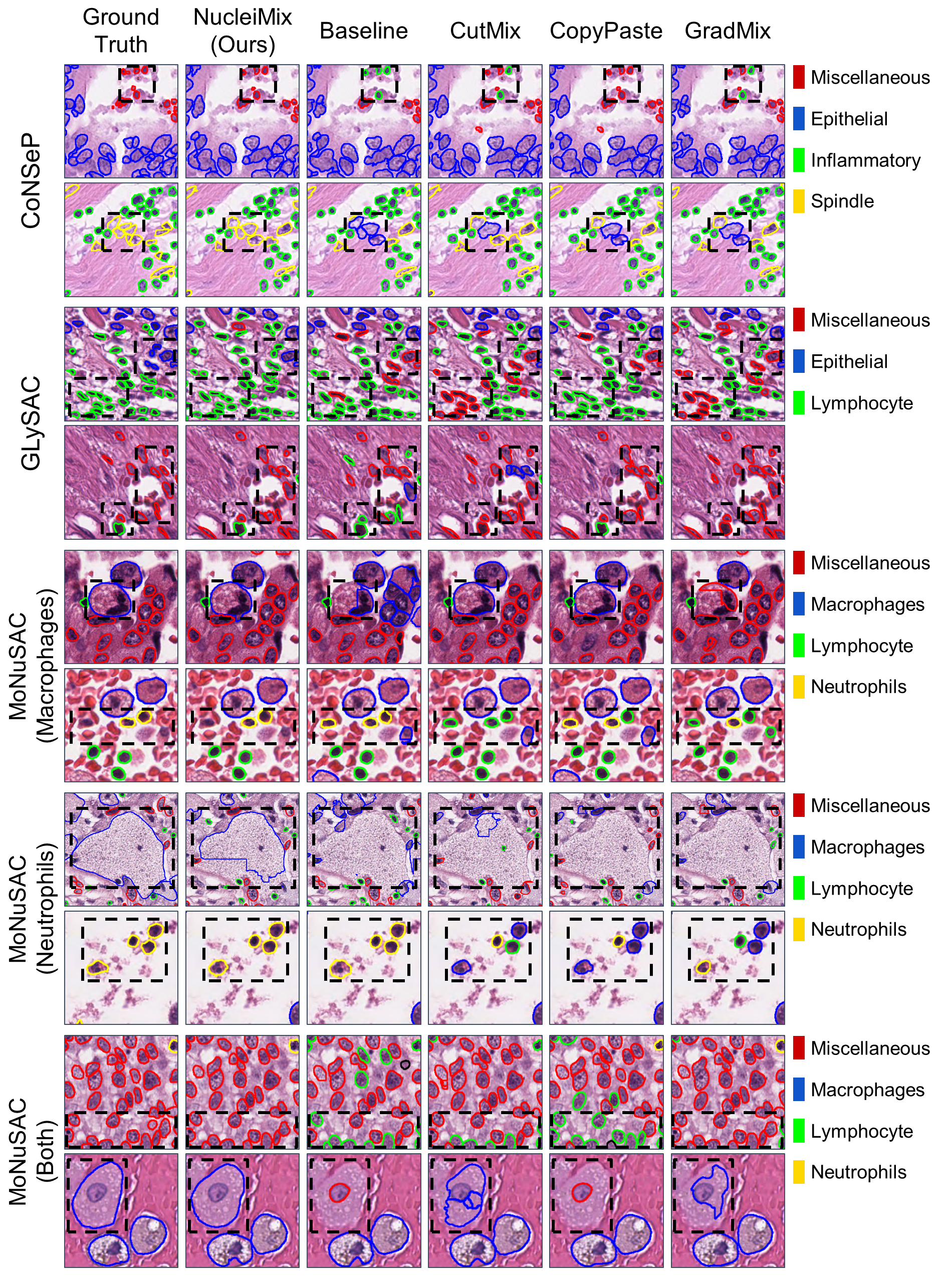} 
    \caption{Visual assessment of augmentation methods using SONNET. Different nuclear types are contoured in different colors, and representative regions are highlighted in dashed black boxes.}
    \label{fig:visual_sonnet}
\end{figure*}

\begin{figure*}[!]
    \centering
    \includegraphics[width=0.99\linewidth] {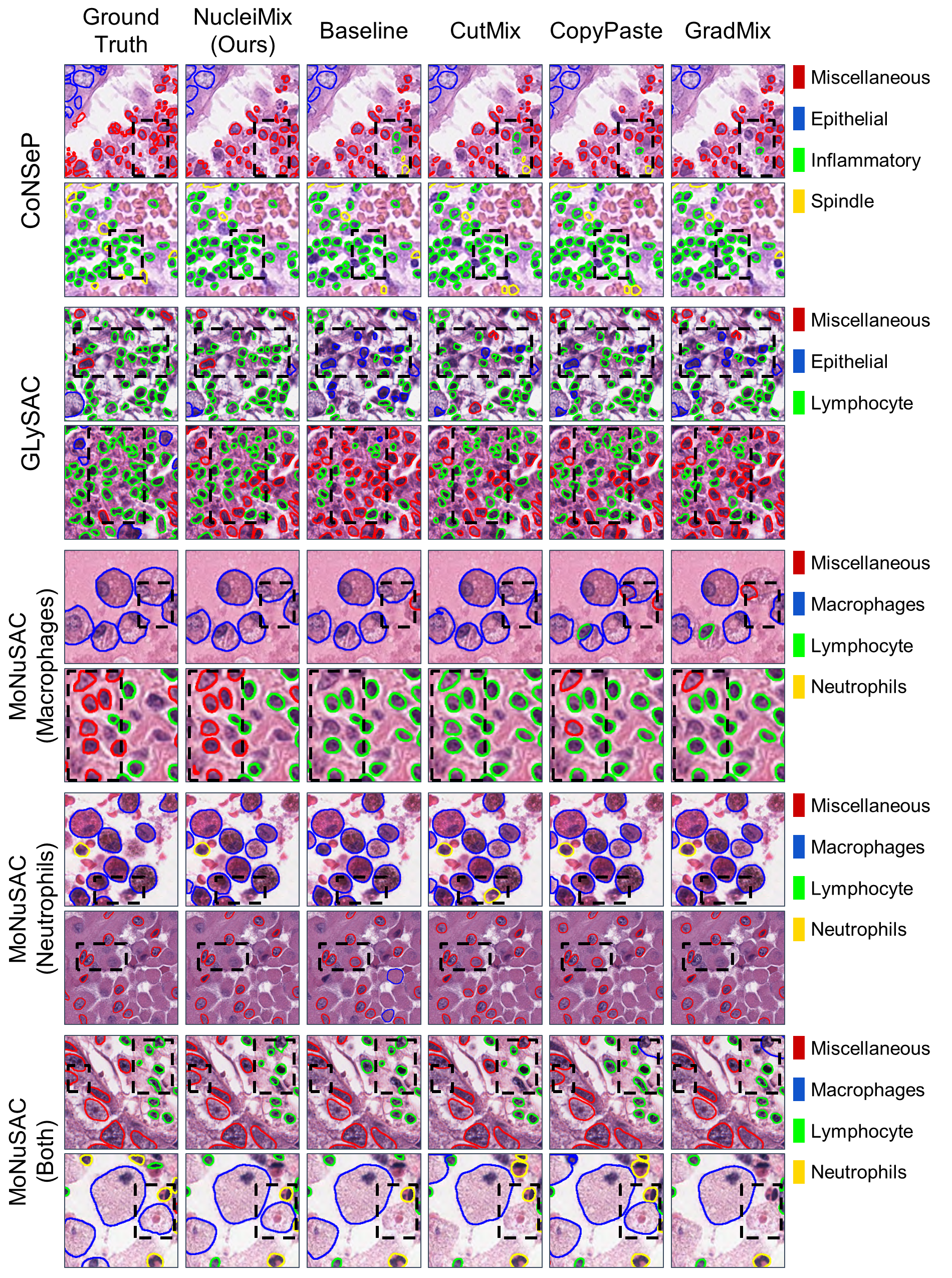}
    \caption{Visual assessment of augmentation methods using PointoNu. Different nuclear types are contoured in different colors, and representative regions are highlighted in dashed black boxes.}
    \label{fig:visual_pointnu}
\end{figure*}

We visually compared NucleiMix with three other methods when a new rare-type nucleus replaces a major-type nucleus. As shown in Fig.~\ref{fig:comparisons}, NucleiMix was capable of generating realistic images for both isolated and touched nuclei, while other methods often produced artifacts or implausible nuclei.
We also visually investigated the results of nuclei segmentation and classification using NucleiMix, comparing it with the baseline and three other augmentation methods. 
Fig.~\ref{fig:visual_sonnet} and Fig.~\ref{fig:visual_pointnu} present qualitative comparisons of nuclei segmentation and classification results for the SONNET and PointNu models, evaluated under baseline and different augmentation method settings. We selected representative regions (highlighted with dashed black boxes) that include multiple nuclei to assess how well each setting preserves boundary accuracy and classification consistency. These visualizations serve to complement aggregate metrics (e.g., DICE, AJI, PQ, DQ, SQ) by providing targeted insights into specific strengths and failure cases. As shown in both figures, NucleiMix consistently outperforms competing methods across different datasets and nuclei types. In contrast, baseline and alternative methods are less accurate by exhibiting errors, including missed detections (e.g., second row of CoNSeP in Fig.~\ref{fig:visual_sonnet}), false positives (e.g., second row of MoNuSAC (Macrophages) in Fig.~\ref{fig:visual_pointnu}), under-segmentation (e.g., first row of MoNuSAC (Macrophages) in Fig.~\ref{fig:visual_sonnet}), and misclassification of nuclei types despite correct detection (e.g., second row of MoNuSAC (Macrophages), second row in Fig.~\ref{fig:visual_sonnet}). These examples highlight NucleiMix’s effectiveness in enhancing both segmentation accuracy and nuclei-type classification reliability.

Regardless of the backbone models, NucleiMix consistently provided improved segmentation and classification results. 
For CoNSeP and GLySAC, both NucleiMix and other augmentation methods generally produced segmentation results similar to the ground truth. However, the other augmentation methods often failed to segment rare-type nuclei or over-segmented the nuclei. 
Similar observations were made in MoNuSAC. Specifically, for macrophages, the other augmentation methods struggled to segment them properly, likely due to their large size, resulting in a tendency to either segment and classify smaller regions or miss them entirely.

\begin{table*}[htbp]
\caption{Results of Nuclei Segmentation  (k = 600)}
\label{tab:seg}
\begin{center}    
\resizebox{\linewidth}{!}{
\begin{tabular}{cccccccccccccc} 
\hline
\rule[-1ex]{0pt}{3.5ex} Model & Augmentation  & DICE  & \(\uparrow\) (\%)& AJI & \(\uparrow\) (\%)  & DQ & \(\uparrow\) (\%)  & SQ  & \(\uparrow\) (\%) & PQ & \(\uparrow\) (\%)  & b-IoU & \(\uparrow\) (\%)   \\ 
\hline
\rule[-1ex]{0pt}{3.5ex} \multirow{6}{6em}{SONNET\\(CoNSeP: Miscellaneous)} & Baseline  & 0.830$\pm$0.001 && 0.479$\pm$0.009 && 0.632$\pm$0.006 && 0.758$\pm$0.002 && 0.480$\pm$0.006 && 0.717±0.001 & \\
\cline{2-14}
\rule[-1ex]{0pt}{3.5ex}  & NucleiMix (Ours) & $\mathbf{0.832\pm0.001}$ &$\mathbf{0.24}$& $\mathbf{0.498\pm0.006}$ &$\mathbf{3.97}$& $\mathbf{0.651\pm0.005}$ &$\mathbf{3.01}$& $\mathbf{0.763\pm0.002}$ &$\mathbf{0.66}$& $\mathbf{0.498\pm0.006}$ &$\mathbf{3.75}$ & $\mathbf{0.720\pm0.003}$ &$\mathbf{0.40}$  \\

\rule[-1ex]{0pt}{3.5ex}  & CutMix & 0.829$\pm$0.001 &-0.12& 0.488$\pm$0.003 &1.88& 0.641$\pm$0.004 &1.42& 0.760$\pm$0.002 &0.26& 0.488$\pm$0.004 &1.67& 0.716±0.003 &-0.03    \\

\rule[-1ex]{0pt}{3.5ex}  & CopyPaste & 0.829$\pm$0.001 &-0.12& 0.481$\pm$0.008 &0.42& 0.637$\pm$0.003 &0.79& 0.762$\pm$0.001 &0.53& 0.486$\pm$0.003 &1.25& 0.718±0.001 &0.16  \\

\rule[-1ex]{0pt}{3.5ex}  & GradMix & 0.829$\pm$0.002 &-0.12& 0.479$\pm$0.005 &0.00& 0.636$\pm$0.006 &0.63& 0.761$\pm$0.002 &0.40& 0.485$\pm$0.005 &1.04& 0.717±0.002 &0.10   \\
\hline

\rule[-1ex]{0pt}{3.5ex} \multirow{6}{6em}{PointNu\\(CoNSeP: Miscellaneous)} 
& Baseline  & 0.743$\pm$0.005 && 0.491$\pm$0.006 && 0.665$\pm$0.001 && 0.757$\pm$0.002 && 0.504$\pm$0.001 && 0.715±0.002 & \\
\cline{2-14}
\rule[-1ex]{0pt}{3.5ex}  & NucleiMix (Ours) & $\mathbf{0.784\pm0.004}$ &$\mathbf{5.52}$& $\mathbf{0.525\pm0.001}$ &$\mathbf{6.92}$& $\mathbf{0.685\pm0.004}$ &$\mathbf{3.01}$& 0.757$\pm$0.002 &0.00& $\mathbf{0.520\pm0.005}$ & $\mathbf{3.17}$ & $\mathbf{0.717\pm0.002}$ &$\mathbf{0.27}$ \\

\rule[-1ex]{0pt}{3.5ex}  & CutMix & 0.778$\pm$0.010 &4.71& 0.521$\pm$0.008 &6.11& 0.681$\pm$0.009 &2.41& 0.754$\pm$0.002 &0.00& 0.515$\pm$0.008 &3.17& 0.715±0.001 &-0.07  \\

\rule[-1ex]{0pt}{3.5ex}  & CopyPaste & 0.778$\pm$0.002 &4.71& 0.521$\pm$0.002 &6.11& 0.682$\pm$0.007 &2.56& $\mathbf{0.758\pm0.004}$ &$\mathbf{0.13}$& 0.518$\pm$0.008 &2.78& 0.716±0.002 &0.10   \\

\rule[-1ex]{0pt}{3.5ex}  & GradMix & 0.777$\pm$0.017 &4.58& 0.522$\pm$0.016 &6.31& 0.679$\pm$0.017 &2.11& 0.755$\pm$0.005 &-0.26& 0.513$\pm$0.016 &1.79& 0.716±0.001 &0.14   \\
\hline

\rule[-1ex]{0pt}{3.5ex} \multirow{6}{6em}{SONNET\\(GLySAC: Miscellaneous)} & Baseline  & 0.822$\pm$0.002 && 0.627$\pm$0.004 && 0.786$\pm$0.005 && 0.782$\pm$0.002 && 0.617$\pm$0.004 && 0.747±0.003 & \\
\cline{2-14}
\rule[-1ex]{0pt}{3.5ex}  & NucleiMix (Ours) & 0.822$\pm$0.000 &0.00& $\mathbf{0.631\pm0.001}$ &$\mathbf{0.64}$& 0.786$\pm$0.004 &0.00& $\mathbf{0.784\pm0.002}$ &$\mathbf{0.26}$& $\mathbf{0.619\pm0.002}$ &$\mathbf{0.32}$& $\mathbf{0.749\pm0.003}$ &$\mathbf{0.33}$   \\

\rule[-1ex]{0pt}{3.5ex}  & CutMix & 0.820$\pm$0.002 &-0.24& 0.628$\pm$0.002 &0.16& 0.786$\pm$0.004 &0.00& 0.783$\pm$0.001 &0.13& 0.618$\pm$0.006 &0.16& 0.748±0.001 &0.12    \\

\rule[-1ex]{0pt}{3.5ex}  & CopyPaste & 0.821$\pm$0.002 &-0.12& 0.629$\pm$0.003 &0.32& $\mathbf{0.787\pm0.003}$ &$\mathbf{0.13}$& 0.782$\pm$0.13 &0.00& 0.618$\pm$0.005 &0.16& 0.746±0.004 &-0.06  \\

\rule[-1ex]{0pt}{3.5ex}  & GradMix & $\mathbf{0.823\pm0.001}$ &$\mathbf{0.12}$& 0.629$\pm$0.004 &0.32& 0.784$\pm$0.003 &-0.25& 0.783$\pm$0.001 &0.13& 0.617$\pm$0.001 &0.00& 0.748±0.001 &0.14   \\
\hline

\rule[-1ex]{0pt}{3.5ex} \multirow{6}{6em}{PointNu\\(GLySAC: Miscellaneous)} & Baseline  & 0.808$\pm$0.001 && 0.629$\pm$0.001 && 0.800$\pm$0.001 && $\mathbf{0.786\pm0.000}$ && 0.630$\pm$0.001 && $\mathbf{0.751\pm0.002}$ & \\
\cline{2-14}
\rule[-1ex]{0pt}{3.5ex}  & NucleiMix (Ours) & 0.808$\pm$0.001 &0.00& 0.629$\pm$0.001 &0.00& $\mathbf{0.802\pm0.001}$ &$\mathbf{0.25}$& 0.786$\pm$0.001 &-0.13& $\mathbf{0.631\pm0.000}$ &$\mathbf{0.16}$& 0.750±0.001 &-0.08  \\

\rule[-1ex]{0pt}{3.5ex}  & CutMix & 0.811$\pm$0.000 &0.37& $\mathbf{0.630\pm0.001}$ &$\mathbf{0.16}$& 0.798$\pm$0.002 &-0.25& 0.781$\pm$0.001 &-0.64& 0.625$\pm$0.002 &-0.79& 0.747±0.001 &-0.42    \\

\rule[-1ex]{0pt}{3.5ex}  & CopyPaste & $\mathbf{0.813\pm0.001}$ &$\mathbf{0.62}$& $\mathbf{0.630\pm0.001}$ &$\mathbf{0.16}$& 0.799$\pm$0.002 &-0.13& 0.782$\pm$0.001 &-0.51& 0.626$\pm$0.002 &-0.63& 0.746±0.001 &-0.61  \\

\rule[-1ex]{0pt}{3.5ex}  & GradMix & 0.811$\pm$0.004 &0.37& 0.630$\pm$0.005 &0.16& 0.801$\pm$0.003 &0.13& 0.783$\pm$0.001 &-0.38& 0.629$\pm$0.004 &-0.16& 0.748±0.002 &-0.30   \\
\hline

\rule[-1ex]{0pt}{3.5ex} \multirow{6}{6em}{SONNET\\(MoNuSAC: Macrophages)} & Baseline  & 0.774$\pm$0.006 && 0.624$\pm$0.006 && 0.788$\pm$0.007 && 0.797$\pm$0.003 && 0.643$\pm$0.004 && 0.463±0.001 & \\
\cline{2-14}
\rule[-1ex]{0pt}{3.5ex}  & NucleiMix (Ours) & $\mathbf{0.782\pm0.001}$ &$\mathbf{1.03}$& $\mathbf{0.630\pm0.004}$ &$\mathbf{0.96}$& $\mathbf{0.795\pm0.005}$ &$\mathbf{0.89}$& 0.816$\pm$0.003 &2.38& $\mathbf{0.650\pm0.004}$ &$\mathbf{1.09}$& $\mathbf{0.469\pm0.005}$ &$\mathbf{1.17}$   \\

\rule[-1ex]{0pt}{3.5ex}  & CutMix & 0.767$\pm$0.003 &-0.90& 0.612$\pm$0.002 &-1.92& 0.783$\pm$0.004 &-0.63& 0.805$\pm$0.010 &1.00& 0.637$\pm$0.003 &-0.93& 0.463±0.009 &0.03    \\

\rule[-1ex]{0pt}{3.5ex}  & CopyPaste & 0.771$\pm$0.005 &-0.39& 0.616$\pm$0.006 &-1.28& 0.785$\pm$0.006 &-0.38& 0.808$\pm$0.011 &1.38& 0.642$\pm$0.005 &-0.16& 0.466±0.008 &0.67  \\

\rule[-1ex]{0pt}{3.5ex}  & GradMix & 0.772$\pm$0.002 &-0.26& 0.622$\pm$0.000 &-0.32& 0.792$\pm$0.002 &0.51& $\mathbf{0.818\pm0.002}$ &$\mathbf{2.63}$& 0.648$\pm$0.001 &0.78& 0.468±0.002 &1.00   \\
\hline

\rule[-1ex]{0pt}{3.5ex} \multirow{6}{6em}{PointNu\\(MoNuSAC: Macrophages)} & Baseline  & 0.742$\pm$0.010 && 0.600$\pm$0.009 && 0.777$\pm$0.009 && 0.797$\pm$0.005 && 0.635$\pm$0.008 && 0.488±0.005 & \\
\cline{2-14}
\rule[-1ex]{0pt}{3.5ex}  & NucleiMix (Ours) & $\mathbf{0.756\pm0.004}$ &$\mathbf{1.89}$& $\mathbf{0.614\pm0.004}$ &$\mathbf{2.33}$& $\mathbf{0.790\pm0.006}$ &$\mathbf{1.67}$& 0.799$\pm$0.004 &0.25& $\mathbf{0.647\pm0.002}$ &$\mathbf{1.89}$& $\mathbf{0.494\pm0.002}$ &$\mathbf{1.27}$   \\

\rule[-1ex]{0pt}{3.5ex}  & CutMix & 0.747$\pm$0.002 &0.67& 0.603$\pm$0.002 &0.50& 0.778$\pm$0.003 &0.13& 0.798$\pm$0.002 &0.13& 0.636$\pm$0.003 &0.16& 0.490±0.006 &0.43    \\

\rule[-1ex]{0pt}{3.5ex}  & CopyPaste & 0.750$\pm$0.009 &1.08& 0.601$\pm$0.013 &0.17& 0.778$\pm$0.008 &0.13& 0.796$\pm$0.004 &-0.13& 0.634$\pm$0.009 &-0.16& 0.487±0.003 &-0.14  \\

\rule[-1ex]{0pt}{3.5ex}  & GradMix & 0.755$\pm$0.002 &1.75& 0.612$\pm$0.008 &2.00& 0.787$\pm$0.005 &1.29& $\mathbf{0.799\pm0.001}$ &$\mathbf{0.25}$& 0.644$\pm$0.003 &1.42& 0.492±0.007 &0.85   \\
\hline

\rule[-1ex]{0pt}{3.5ex} \multirow{6}{6em}{SONNET\\(MoNuSAC: Neutrophils)} & Baseline  & 0.774$\pm$0.006 && 0.624$\pm$0.006 && 0.788$\pm$0.007 && 0.797$\pm$0.003 && 0.643$\pm$0.004 && 0.463±0.001 & \\
\cline{2-14}
\rule[-1ex]{0pt}{3.5ex}  & NucleiMix (Ours) & $\mathbf{0.778\pm0.002}$ &$\mathbf{0.52}$& 0.627$\pm$0.001 &0.48& 0.792$\pm$0.001 &0.51& $\mathbf{0.819\pm0.001}$ &$\mathbf{2.76}$& $\mathbf{0.650\pm0.001}$ &$\mathbf{1.09}$& $\mathbf{0.473\pm0.002}$ &$\mathbf{2.06}$   \\

\rule[-1ex]{0pt}{3.5ex}  & CutMix & 0.777$\pm$0.004 &0.39& 0.629$\pm$0.002 &0.80& 0.791$\pm$0.005 &0.38& 0.800$\pm$0.009 &0.38& 0.644$\pm$0.003 &0.16& 0.463±0.006 &-0.05    \\

\rule[-1ex]{0pt}{3.5ex}  & CopyPaste & $\mathbf{0.778\pm0.002}$ &$\mathbf{0.52}$& 0.630$\pm$0.003 &0.96& 0.794$\pm$0.004 &0.76& 0.804$\pm$0.012 &0.88& 0.649$\pm$0.005 &0.93& 0.468$\pm$0.004 &1.05  \\

\rule[-1ex]{0pt}{3.5ex}  & GradMix & 0.776$\pm$0.001 &0.26& $\mathbf{0.630\pm0.002}$ &$\mathbf{0.96}$& $\mathbf{0.796\pm0.004}$ &$\mathbf{1.02}$& 0.795$\pm$0.002 &-0.25& 0.648$\pm$0.004 &0.783& 0.464±0.006 &0.11   \\
\hline

\rule[-1ex]{0pt}{3.5ex} \multirow{6}{6em}{PointNu\\(MoNuSAC: Neutrophils)} & Baseline  & 0.742$\pm$0.010 && 0.600$\pm$0.009 && 0.777$\pm$0.009 && 0.797$\pm$0.005 && 0.635$\pm$0.008 && 0.488±0.005 & \\
\cline{2-14}
\rule[-1ex]{0pt}{3.5ex}  & NucleiMix (Ours) & $\mathbf{0.764\pm0.005}$ &$\mathbf{2.97}$& $\mathbf{0.624\pm0.007}$ &$\mathbf{4.00}$& $\mathbf{0.797\pm0.003}$ &$\mathbf{2.53}$& 0.799$\pm$0.002 &0.33& $\mathbf{0.653\pm0.004}$ &$\mathbf{2.89}$& 0.492±0.000 &0.73   \\

\rule[-1ex]{0pt}{3.5ex}  & CutMix & 0.755$\pm$0.011 &1.84& 0.610$\pm$0.011 &1.67& 0.788$\pm$0.007 &1.42& 0.796$\pm$0.003 &-0.08& 0.643$\pm$0.007 &1.26& 0.486±0.009 &-0.45    \\

\rule[-1ex]{0pt}{3.5ex}  & CopyPaste & 0.760$\pm$0.007 &2.43& 0.616$\pm$0.008 &2.78& 0.789$\pm$0.003 &1.50& $\mathbf{0.801\pm0.004}$ &$\mathbf{0.50}$& 0.647$\pm$0.006 &1.94& $\mathbf{0.494\pm0.004}$ &$\mathbf{1.23}$  \\

\rule[-1ex]{0pt}{3.5ex}  & GradMix & 0.756$\pm$0.007 &1.89& 0.612$\pm$0.008 &2.11& 0.784$\pm$0.007 &0.90& 0.797$\pm$0.003 &0.08& 0.640$\pm$0.006 &0.89& 0.488±0.010 &-0.02   \\
\hline

\rule[-1ex]{0pt}{3.5ex} \multirow{6}{6em}{SONNET\\(MoNuSAC: Macrophages and Neutrophils)} & Baseline  & 0.774$\pm$0.006 && 0.624$\pm$0.006 && 0.788$\pm$0.007 && 0.797$\pm$0.003 && 0.643$\pm$0.004 && 0.463±0.001 & \\
\cline{2-14}
\rule[-1ex]{0pt}{3.5ex}  & NucleiMix (Ours) & $\mathbf{0.781\pm0.005}$ &$\mathbf{0.90}$& $\mathbf{0.632\pm0.007}$&$\mathbf{1.28}$& $\mathbf{0.794\pm0.008}$ &$\mathbf{0.76}$& 0.809$\pm$0.006 &1.51& $\mathbf{0.649\pm0.004}$ &$\mathbf{0.93}$& $\mathbf{0.470\pm0.004}$ &$\mathbf{1.35}$   \\

\rule[-1ex]{0pt}{3.5ex}  & CutMix & 0.772$\pm$0.005 &-0.26& 0.622$\pm$0.007 &-0.32& 0.785$\pm$0.005 &-0.38& 0.803$\pm$0.008 &0.75& 0.641$\pm$0.004 &-0.31& 0.467±0.003 &0.78    \\

\rule[-1ex]{0pt}{3.5ex}  & CopyPaste & 0.774$\pm$0.008 &0.00& 0.624$\pm$0.009 &0.00& 0.790$\pm$0.010 &0.25& $\mathbf{0.810\pm0.008}$ &$\mathbf{1.63}$& 0.646$\pm$0.008 &0.47& 0.469±0.001 &1.28  \\

\rule[-1ex]{0pt}{3.5ex}  & GradMix & 0.772$\pm$0.004 &-0.26& 0.622$\pm$0.004 &-0.32& 0.792$\pm$0.001 &0.51& 0.808$\pm$0.008 &1.38& 0.647$\pm$0.002 &0.62& 0.466±0.004 &0.69   \\
\hline

\rule[-1ex]{0pt}{3.5ex} \multirow{6}{6em}{PointNu\\(MoNuSAC: Macrophages and Neutrophils)} & Baseline  & 0.742$\pm$0.010 && 0.600$\pm$0.009 && 0.777$\pm$0.009 && 0.797$\pm$0.005 && 0.635$\pm$0.008 && 0.488±0.005 & \\
\cline{2-14}
\rule[-1ex]{0pt}{3.5ex}  & NucleiMix (Ours) & $\mathbf{0.757\pm0.000}$ &$\mathbf{2.02}$& $\mathbf{0.616\pm0.002}$ &$\mathbf{2.67}$& $\mathbf{0.795\pm0.002}$ &$\mathbf{2.32}$& 0.796$\pm$0.004 &-0.13& $\mathbf{0.649\pm0.002}$ &$\mathbf{2.20}$& 0.492±0.006 &0.80   \\

\rule[-1ex]{0pt}{3.5ex}  & CutMix & 0.749$\pm$0.011 &0.94& 0.605$\pm$0.011 &0.83& 0.780$\pm$0.004 &0.39& 0.800$\pm$0.001 &0.38& 0.639$\pm$0.004 &0.63& 0.493±0.005 &0.91    \\

\rule[-1ex]{0pt}{3.5ex}  & CopyPaste & 0.757$\pm$0.004 &2.02& 0.609$\pm$0.004 &1.50& 0.785$\pm$0.004 &1.03& $\mathbf{0.803\pm0.009}$ &$\mathbf{0.75}$& 0.642$\pm$0.004 &1.10& $\mathbf{0.495\pm0.002}$ &$\mathbf{1.38}$  \\

\rule[-1ex]{0pt}{3.5ex}  & GradMix & 0.748$\pm$0.009 &0.81& 0.603$\pm$0.011 &0.50& 0.786$\pm$0.006 &1.16& 0.796$\pm$0.003 &-0.13& 0.641$\pm$0.007 &0.94& 0.490±0.003 &0.47   \\
\hline
\end{tabular}
}
\end{center}
\end{table*}

\begin{table*}[htbp]
\caption{Results of Nuclei Classification (k = 600)}
\label{tab:cls}
\begin{center}    
\resizebox{\linewidth}{!}{
\begin{tabular}{cccccccccccccc} 
\hline
\rule[-1ex]{0pt}{3.5ex} Model & Augmentation  & $F_d$  & \(\uparrow\) (\%)& $Acc$ & \(\uparrow\) (\%)  & $F^M_c$ & \(\uparrow\) (\%)  & $F^I_c$  & \(\uparrow\) (\%) & $F^E_c$ & \(\uparrow\)(\%)   & $F^S_c$  & \(\uparrow\) (\%)  \\ 
\hline
\rule[-1ex]{0pt}{3.5ex} \multirow{6}{6em}{SONNET\\(CoNSeP: Miscellaneous)} & Baseline  & 0.741$\pm$0.003 && 0.852$\pm$0.003 && 0.352$\pm$0.051 && 0.595$\pm$0.006 && 0.607$\pm$0.010 && 0.555$\pm$0.008 &\\
\cline{2-14}
\rule[-1ex]{0pt}{3.5ex}  & NucleiMix (Ours) & $\mathbf{0.742\pm0.002}$ &$\mathbf{0.13}$& $\mathbf{0.855\pm0.002}$ &$\mathbf{0.35}$& $\mathbf{0.410\pm0.041}$ &$\mathbf{16.48}$& $\mathbf{0.598\pm0.016}$ &$\mathbf{0.50}$& $\mathbf{0.608\pm0.004}$ &$\mathbf{0.16}$& $\mathbf{0.555\pm0.005}$ &$\mathbf{0.00}$ \\
\rule[-1ex]{0pt}{3.5ex}  & CutMix & 0.735$\pm$0.001 &-0.81& 0.852$\pm$0.002 &0.00& 0.348$\pm$0.019 &-1.14& 0.597$\pm$0.015 &0.34& 0.602$\pm$0.002 &-0.82& 0.551$\pm$0.004  &-0.72 \\
\rule[-1ex]{0pt}{3.5ex}  & CopyPaste & 0.733$\pm$0.007 &-1.08& 0.854$\pm$0.001 &0.23& 0.354$\pm$0.018 &0.57& 0.582$\pm$0.010 &-2.18& 0.606$\pm$0.012 &-0.16& 0.555$\pm$0.006 &0.00  \\
\rule[-1ex]{0pt}{3.5ex}  & GradMix & 0.742$\pm$0.004 &0.13& 0.846$\pm$0.003 &-0.70& 0.380$\pm$0.022 &7.95& 0.593$\pm$0.009 &-0.34& 0.597$\pm$0.004 &-1.65& 0.545$\pm$0.009 &-1.80  \\
\hline
\rule[-1ex]{0pt}{3.5ex} \multirow{6}{6em}{PointNu\\(CoNSeP: Miscellaneous)} & Baseline  & 0.724$\pm$0.003 && 0.888$\pm$0.005 && 0.544$\pm$0.021 && 0.661$\pm$0.020 && 0.620$\pm$0.014 && 0.542$\pm$0.014 &\\
\cline{2-14}
\rule[-1ex]{0pt}{3.5ex}  & NucleiMix (Ours) & $\mathbf{0.744\pm0.004}$ &$\mathbf{2.76}$& $\mathbf{0.897\pm0.007}$ &$\mathbf{1.01}$& $\mathbf{0.600\pm0.005}$ &$\mathbf{10.29}$& $\mathbf{0.680\pm0.006}$ &$\mathbf{2.87}$& $\mathbf{0.651\pm0.009}$ &$\mathbf{5.00}$& $\mathbf{0.558\pm0.019}$ &$\mathbf{2.95}$  \\
\rule[-1ex]{0pt}{3.5ex}  & CutMix & 0.738$\pm$0.003 &1.93& 0.890$\pm$0.007 &0.23& 0.590$\pm$0.015 &8.46& 0.665$\pm$0.014 &0.61& 0.644$\pm$0.011 &3.87& 0.542$\pm$0.012  &0.00 \\
\rule[-1ex]{0pt}{3.5ex}  & CopyPaste & 0.741$\pm$0.005 &2.35& 0.895$\pm$0.003 &0.79& 0.576$\pm$0.005 &5.88& 0.666$\pm$0.008 &0.76& 0.650$\pm$0.008 &4.84& 0.557$\pm$0.004  &2.77 \\
\rule[-1ex]{0pt}{3.5ex}  & GradMix & 0.730$\pm$0.024 &0.83& 0.890$\pm$0.005 &0.23& 0.589$\pm$0.013 &8.27& 0.637$\pm$0.021 &-3.63& 0.632$\pm$0.022 &1.94& 0.556$\pm$0.001 &2.58  \\
\hline

\rule[-1ex]{0pt}{3.5ex} Model & Augmentation  & $F_d$  & \(\uparrow\) (\%)& $Acc$ & \(\uparrow\) (\%)  & $F^M_c$ & \(\uparrow\) (\%)  & $F^L_c$  & \(\uparrow\) (\%) & $F^E_c$ & \(\uparrow\)(\%)  \\ 
\hline
\rule[-1ex]{0pt}{3.5ex} \multirow{6}{6em}{SONNET\\(GLySAC: Miscellaneous)} & Baseline  & 0.838$\pm$0.003 && 0.698$\pm$0.004 && 0.287$\pm$0.05 && 0.520$\pm$0.007 && 0.540$\pm$0.005 &\\
\cline{2-12}
\rule[-1ex]{0pt}{3.5ex}  & NucleiMix (Ours) & $\mathbf{0.843\pm0.004}$ &$\mathbf{0.60}$& $\mathbf{0.706\pm0.005}$ &$\mathbf{1.15}$& $\mathbf{0.310\pm0.011}$ &$\mathbf{8.01}$& $\mathbf{0.531\pm0.013}$ &$\mathbf{2.12}$& $\mathbf{0.545\pm0.003 }$  &$\mathbf{0.93}$\\
\rule[-1ex]{0pt}{3.5ex}  & CutMix & 0.838$\pm$0.002 &0.00& 0.704$\pm$0.003 &0.86& 0.300$\pm$0.008 &4.53& 0.527$\pm$0.009 &1.35& 0.541$\pm$0.006 &0.19 \\
\rule[-1ex]{0pt}{3.5ex}  & CopyPaste & 0.838$\pm$0.001 &0.00& 0.705$\pm$0.004 &1.00& 0.302$\pm$0.006 &5.23& 0.520$\pm$0.008 &0.00& 0.544$\pm$0.010  &0.74 \\
\rule[-1ex]{0pt}{3.5ex}  & GradMix & 0.834$\pm$0.012 &-0.48& 0.695$\pm$0.015 &-0.43& 0.296$\pm$0.005 &3.14& 0.510$\pm$0.033 &-1.92& 0.544$\pm$0.004  &0.74\\
\hline
\rule[-1ex]{0pt}{3.5ex} \multirow{6}{6em}{PointNu\\(GLySAC: Miscellaneous)} & Baseline  & 0.824$\pm$0.002 && 0.715$\pm$0.010 && 0.294$\pm$0.010 && 0.518$\pm$0.019 && 0.553$\pm$0.004 &\\
\cline{2-12}
\rule[-1ex]{0pt}{3.5ex}  & NucleiMix (Ours) & 0.832$\pm$0.005 &0.97& $\mathbf{0.720\pm0.006}$ &$\mathbf{0.70}$& $\mathbf{0.318\pm0.004}$ &$\mathbf{8.16}$& $\mathbf{0.525\pm0.005}$ &$\mathbf{1.35}$& $\mathbf{0.563\pm0.010}$ &$\mathbf{1.81}$  \\
\rule[-1ex]{0pt}{3.5ex}  & CutMix & 0.829$\pm$0.004 &0.61& 0.716$\pm$0.008 &0.14& 0.303$\pm$0.008 &3.06& 0.516$\pm$0.011 &-0.39& 0.557$\pm$0.013  &0.72 \\
\rule[-1ex]{0pt}{3.5ex}  & CopyPaste & 0.832$\pm$0.005  &0.97& 0.711$\pm$0.011 &-0.56& 0.290$\pm$0.010 &-1.36& 0.520$\pm$0.021 &0.39& 0.559$\pm$0.005  &1.08 \\
\rule[-1ex]{0pt}{3.5ex}  & GradMix & $\mathbf{0.832\pm0.004}$ &$\mathbf{0.97}$& 0.717$\pm$0.007 &0.28& 0.297$\pm$0.014 &1.02& 0.520$\pm$0.019 &0.39& 0.553$\pm$0.008  &0.00 \\
\hline

\rule[-1ex]{0pt}{3.5ex} Model & Augmentation  & $F_d$  &\(\uparrow\) (\%) & $Acc$  &\(\uparrow\) (\%) & $F^E_c$ &\(\uparrow\) (\%) & $F^L_c$  &\(\uparrow\) (\%) & $F^M_c$  &\(\uparrow\) (\%) & $F^N_c$  &\(\uparrow\) (\%) \\ 
\hline

\rule[-1ex]{0pt}{3.5ex} \multirow{6}{6em}{SONNET\\(MoNuSAC: Macrophages)} & Baseline  & 0.830$\pm$0.001 && 0.939$\pm$0.004 && 0.709$\pm$0.005 && 0.786$\pm$0.008 && 0.511$\pm$0.017 && 0.514$\pm$0.028 &\\
\cline{2-14}
\rule[-1ex]{0pt}{3.5ex}  & NucleiMix (Ours) & $\mathbf{0.835\pm0.002}$ &$\mathbf{0.60}$& $\mathbf{0.952\pm0.007}$ &$\mathbf{1.38}$& $\mathbf{0.746\pm0.010}$ &$\mathbf{5.22}$& $\mathbf{0.801\pm0.006}$ &$\mathbf{1.91}$& $\mathbf{0.532\pm0.009}$ &$\mathbf{4.11}$& $\mathbf{0.535\pm0.019}$  &$\mathbf{4.09}$ \\
\rule[-1ex]{0pt}{3.5ex}  & CutMix & 0.830$\pm$0.008 &0.00& 0.940$\pm$0.007 &0.11& 0.710$\pm$0.003 &0.14& 0.790$\pm$0.008 &0.51& 0.516$\pm$0.005 &0.98& 0.485$\pm$0.016  &-5.64\\
\rule[-1ex]{0pt}{3.5ex}  & CopyPaste & 0.830$\pm$0.005 &0.00& 0.937$\pm$0.004 &-0.21& 0.718$\pm$0.005 &1.27& 0.778$\pm$0.001 &-1.02& 0.514$\pm$0.022 &0.59& 0.532$\pm$0.006  &3.50 \\
\rule[-1ex]{0pt}{3.5ex}  & GradMix & 0.831$\pm$0.001 &0.12& 0.947$\pm$0.000 &0.85& 0.730$\pm$0.000 &2.96& 0.800$\pm$0.001 &1.78& 0.513$\pm$0.022 &0.39& 0.494$\pm$0.004  &-3.89\\
\hline
\rule[-1ex]{0pt}{3.5ex} \multirow{6}{6em}{PointNu\\(MoNuSAC: Macrophages)} & Baseline  & 0.835$\pm$0.001 && 0.975$\pm$0.001 && 0.786$\pm$0.004 && 0.828$\pm$0.007 && 0.501$\pm$0.018 && 0.508$\pm$0.036 \\
\cline{2-14}
\rule[-1ex]{0pt}{3.5ex}  & NucleiMix (Ours) & $\mathbf{0.840\pm0.006}$ &$\mathbf{0.60}$& $\mathbf{0.976\pm0.001}$ &$\mathbf{0.10}$& $\mathbf{0.796\pm0.006}$ &$\mathbf{1.27}$& $\mathbf{0.832\pm0.003}$ &$\mathbf{0.48}$& $\mathbf{0.545\pm0.003}$ &$\mathbf{8.78}$& $\mathbf{0.528\pm0.008}$  &$\mathbf{3.94}$\\
\rule[-1ex]{0pt}{3.5ex}  & CutMix & 0.835$\pm$0.005 &0.00& 0.975$\pm$0.003 &0.00& 0.786$\pm$0.010 &0.00& 0.827$\pm$0.008 &-0.12& 0.507$\pm$0.024 &1.20& 0.500$\pm$0.006  &-1.57\\
\rule[-1ex]{0pt}{3.5ex}  & CopyPaste & 0.831$\pm$0.008 &-0.48& 0.976$\pm$0.001 &0.10& 0.790$\pm$0.002 &0.51& 0.823$\pm$0.013 &-0.60& 0.507$\pm$0.021 &1.20& 0.516$\pm$0.043 &1.57 \\
\rule[-1ex]{0pt}{3.5ex}  & GradMix & 0.838$\pm$0.003 &0.36& 0.975$\pm$0.002 &0.00& 0.792$\pm$0.003 &0.76& 0.829$\pm$0.004 &0.12& 0.531$\pm$0.021 &5.99& 0.492$\pm$0.012  &-3.15 \\
\hline

\rule[-1ex]{0pt}{3.5ex} \multirow{6}{6em}{SONNET\\(MoNuSAC: Neutrophils)} & Baseline  & $\mathbf{0.830\pm0.001}$ && 0.939$\pm$0.004 && 0.709$\pm$0.005 && 0.786$\pm$0.008 && 0.511$\pm$0.017 && $\mathbf{0.514\pm0.028}$ &\\
\cline{2-14}
\rule[-1ex]{0pt}{3.5ex}  & NucleiMix (Ours) & 0.827$\pm$0.003 &-0.36& 0.946$\pm$0.004 &0.75& 0.725$\pm$0.002 &2.26& 0.787$\pm$0.016 &0.13& $\mathbf{0.529\pm0.004}$ &$\mathbf{3.52}$& 0.503$\pm$0.005 &-2.14  \\
\rule[-1ex]{0pt}{3.5ex}  & CutMix & 0.827$\pm$0.003 &-0.36& $\mathbf{0.947\pm0.007}$ &$\mathbf{0.85}$& 0.728$\pm$0.012 &2.68& 0.787$\pm$0.005 &0.13& 0.522$\pm$0.021 &2.15& 0.486$\pm$0.038  &-5.45 \\
\rule[-1ex]{0pt}{3.5ex}  & CopyPaste & 0.830$\pm$0.007 &0.00& 0.945$\pm$0.006 &0.64& 0.725$\pm$0.010 &2.26& 0.794$\pm$0.012 &1.02& 0.524$\pm$0.020 &2.54& 0.490$\pm$0.028  &-4.67 \\
\rule[-1ex]{0pt}{3.5ex}  & GradMix & 0.828$\pm$0.002 &-0.24& 0.945$\pm$0.015 &0.64& $\mathbf{0.729\pm0.011}$ &$\mathbf{2.82}$& $\mathbf{0.795\pm0.011}$ &$\mathbf{1.15}$& 0.526$\pm$0.025 &2.94& 0.495$\pm$0.030  &-3.70 \\
\hline
\rule[-1ex]{0pt}{3.5ex} \multirow{6}{6em}{PointNu\\(MoNuSAC: Neutrophils)} & Baseline  & 0.835$\pm$0.001 && 0.975$\pm$0.001 && 0.786$\pm$0.004 && 0.828$\pm$0.007 && 0.501$\pm$0.018 && 0.508$\pm$0.036 &\\
\cline{2-14}
\rule[-1ex]{0pt}{3.5ex}  & NucleiMix (Ours) & $\mathbf{0.839\pm0.005}$ &$\mathbf{0.48}$& $\mathbf{0.976\pm0.002}$ &$\mathbf{0.10}$& $\mathbf{0.794\pm0.008}$ &$\mathbf{1.02}$& $\mathbf{0.829\pm0.002}$ &$\mathbf{0.12}$& $\mathbf{0.534\pm0.011}$ &$\mathbf{6.59}$& $\mathbf{0.538\pm0.021}$ &$\mathbf{5.91}$ \\
\rule[-1ex]{0pt}{3.5ex}  & CutMix & 0.830$\pm$0.002 &-0.60& 0.973$\pm$0.002 &-0.21& 0.785$\pm$0.002 &-0.13& 0.818$\pm$0.007 &-1.21& 0.512$\pm$0.015 &2.20& 0.524$\pm$0.026  &3.15 \\
\rule[-1ex]{0pt}{3.5ex}  & CopyPaste & 0.835$\pm$0.001 &0.00& 0.975$\pm$0.001 &0.00& 0.788$\pm$0.003 &0.25& 0.826$\pm$0.006 &-0.24& 0.526$\pm$0.001 &4.99& 0.529$\pm$0.014 &4.13  \\
\rule[-1ex]{0pt}{3.5ex}  & GradMix & 0.833$\pm$0.004 &-0.24& 0.975$\pm$0.001 &0.00& 0.789$\pm$0.005 &0.38& 0.822$\pm$0.001 &-0.72& 0.525$\pm$0.018 &4.79& 0.529$\pm$0.024  &4.13\\
\hline

\rule[-1ex]{0pt}{3.5ex} \multirow{6}{6em}{SONNET\\(MoNuSAC: Macrophages and Neutrophils)} & Baseline  & $\mathbf{0.830\pm0.001}$  && 0.939$\pm$0.004 && 0.709$\pm$0.005 && 0.786$\pm$0.008 && 0.511$\pm$0.017 && 0.514$\pm$0.028 &\\
\cline{2-14}
\rule[-1ex]{0pt}{3.5ex}  & NucleiMix (Ours) & $\mathbf{0.830\pm0.001}$ &$\mathbf{0.00}$& $\mathbf{0.954\pm0.003}$ &$\mathbf{1.60}$& $\mathbf{0.740\pm0.018}$ &$\mathbf{4.37}$& $\mathbf{0.802\pm0.007}$ &$\mathbf{2.04}$& $\mathbf{0.548\pm0.014}$ &$\mathbf{7.24}$& $\mathbf{0.538\pm0.006}$ &$\mathbf{4.67}$ \\
\rule[-1ex]{0pt}{3.5ex}  & CutMix & 0.825$\pm$0.005 &-0.60& 0.943$\pm$0.004 &0.43& 0.720$\pm$0.007 &1.55& 0.781$\pm$0.004 &-0.64& 0.521$\pm$0.003 &1.96& 0.511$\pm$0.040 &-0.58  \\
\rule[-1ex]{0pt}{3.5ex}  & CopyPaste & 0.828$\pm$0.002 &-0.24& 0.937$\pm$0.004 &-0.21& 0.704$\pm$0.004 &-0.71& 0.784$\pm$0.005 &-0.25& 0.508$\pm$0.007 &-0.59& 0.513$\pm$0.015 &-0.19 \\
\rule[-1ex]{0pt}{3.5ex}  & GradMix & 0.829$\pm$0.005 &-0.12& 0.951$\pm$0.002 &1.28& 0.738$\pm$0.003 &4.09& 0.798$\pm$0.003 &1.53& 0.526$\pm$0.019 &2.94& 0.478$\pm$0.038 &-7.00  \\
\hline
\rule[-1ex]{0pt}{3.5ex} \multirow{6}{6em}{PointNu\\(MoNuSAC: Macrophages and Neutrophils)} & Baseline  & 0.835$\pm$0.001 && 0.975$\pm$0.001 && 0.786$\pm$0.004 && 0.828$\pm$0.007 && 0.501$\pm$0.018 && 0.508$\pm$0.036 \\
\cline{2-14}
\rule[-1ex]{0pt}{3.5ex}  & NucleiMix (Ours) & $\mathbf{0.844\pm0.002}$ &$\mathbf{1.08}$& 0.974$\pm$0.003 &-0.10& 0.799$\pm$0.008 &1.65& 0.829$\pm$0.007 &0.12& $\mathbf{0.534\pm0.018}$ &$\mathbf{6.59}$& $\mathbf{0.530\pm0.013}$  & $\mathbf{4.33}$ \\
\rule[-1ex]{0pt}{3.5ex}  & CutMix & 0.839$\pm$0.008 &0.48& $\mathbf{0.977\pm0.001}$ &$\mathbf{0.21}$& $\mathbf{0.799\pm0.005}$ &$\mathbf{1.65}$& $\mathbf{0.831\pm0.008}$ &$\mathbf{0.36}$& 0.512$\pm$0.014 &2.20& 0.512$\pm$0.014 &0.79  \\
\rule[-1ex]{0pt}{3.5ex}  & CopyPaste & 0.836$\pm$0.006 &0.12& 0.976$\pm$0.002 &0.10& 0.789$\pm$0.009 &0.38& 0.829$\pm$0.010 &0.12& 0.503$\pm$0.023 &0.40& 0.519$\pm$0.008  &2.17 \\
\rule[-1ex]{0pt}{3.5ex}  & GradMix & 0.836$\pm$0.006 &0.12& 0.975$\pm$0.001 &0.00& 0.793$\pm$0.004 &0.89& 0.825$\pm$0.008 &-0.36& 0.506$\pm$0.012 &1.00& 0.510$\pm$0.011 &0.39 \\
\hline
\end{tabular}
}
\end{center}
\end{table*}

\subsection{CoNSeP: Augmentation of Miscellaneous Nuclei}
Table \ref{tab:seg} shows the results of nuclei segmentation on CoNSeP using SONNET and PointNu with and without synthesized miscellaneous nuclei. For both models, NucleiMix provided a substantial performance gain across five evaluation metrics, achieving the best results except SQ for PointNu. 
In comparison to the baseline, NucleiMix improved SONNET by 0.24\% in DICE, 3.97\% in AJI, 3.01\% in DQ, 0.66\% in SQ, 3.75\% in PQ, and 0.40\% in b-IoU. Similarly, it enhanced PointNu by 5.52\% in DICE, 6.92\% in AJI, 3.01\% in DQ, 3.17\% in PQ, and 0.27\% in b-IoU. The other three augmentation methods generally improved performance over the baseline but were inferior to NucleiMix. 

Table \ref{tab:cls} demonstrates the nuclei classification results on CoNSeP for both models. 
NucleiMix achieved the best performance regardless of evaluation metrics and nuclei classification models. In particular, it substantially improved $F^M_c$ by 16.48\% and 10.29\% for SONNET and PointNu, respectively, as compared to the baseline, indicating its effectiveness in augmenting rare-type nuclei. The effect of the other augmentation methods was disproportionate across different evaluation metrics and models. For instance, GradMix with SONNET and PointNu increased $F^M_c$ by 7.95\% with and 8.27\%, respectively; however $F^I_c$ was dropped by 3.63\% with PointNu, and $F_1$ for all other three types ($F^I_c$, $F^E_c$, and $F^S_c$) decreased with SONNET. Similarly, using SONNET, CutMix, and CopyPaste showed worse performance than the baseline, and with PointNu, they obtained lower improvements compared to NucleMix.

\subsection{GLySAC Dataset: Augmentation of Miscellaneous Nuclei}
The results of nuclei segmentation, both with and without augmented miscellaneous nuclei, are reported in Table \ref{tab:seg}. NucleiMix was generally comparable or superior to the baseline and other three augmentation methods for both SONNET and PointNu. For instance, NucleiMix obtained the highest AJI of 0.631, SQ of 0.784, PQ of 0.619, and b-IoU of 0.749 with SONNET and the best DQ of 0.802 and PQ of 0.631 with PointNu. 
The performance of the other three augmentation methods was inconsistent across evaluation metrics. 
CopyPaste achieved the highest DQ of 0.787 with SONNET and the highest DICE and AJI of 0.813 and 0.630, respectively, with PointNu; however, DICE for SONNET and DQ, SQ, and PQ for PointNu were dropped compared to the baseline. 
GradMix, CutMix, and baseline excelled in only one evaluation metric (GradMix: 0.823 DICE with SONNET; CutMix: 0.630 AJI with PointNu; Baseline: 0.786 SQ with PointNu).

In the nuclei classification on GLySAC (Table \ref{tab:cls}), the superiority of NucleiMix was apparent. NucleiMix outperformed the baseline and other three augmentation methods except $F_d$ for PointNu. The improvement by NucleiMix varied across different types of nuclei, with the greatest improvement observed for miscellaneous nuclei, achieving an increase of 8.01\% with SONNET and 8.16\% and PointNu. Although GradMix showed superior performance in $F_d$ with PointNu, it showed worse performance in all other metrics than NucleiMix. The other two augmentation methods (CutMix and CopyPaste) generally showed improvements over the baseline but exhibited declines in certain metrics.

\subsection{MoNuSAC Dataset: Augmentation of Macrophages, Neutrophils, and Both}
We demonstrate the results of nuclei segmentation on MoNuSAC with augmented macrophages, neutrophils, and both in Table \ref{tab:seg}. 
By augmenting macrophages only, NucleiMix outperformed the baseline across all evaluation metrics, with improvements ranging from 0.89\% to 2.38\%, irrespective of the backbone model. As compared to other augmentation methods, NucleiMix demonstrated superior performance across all evaluation metrics, except SQ with both models. The performance of the other three augmentation methods was inconsistent across both evaluation metrics and backbone models. With SONNET, in particular, they were generally inferior to the baseline.
With augmented neutrophils, NucleiMix outperformed the baseline across all evaluation metrics, with improvements ranging from 0.48\% to 2.76\%. Compared to the other three augmentation methods, NucleiMix generally achieved the highest scores, except for AJI and DQ with SONNET and SQ, b-IoU with PointNu.
In regard to the other three augmentation methods, there was no clear second-best method. For example, with SONNET, GradMix obtained the best AJI and DQ of 0.630 and 0.796, respectively, and CopyPaste demonstrated the highest DICE of 0.778 with SONNET (matching the performance of NucleiMix) and the highest SQ of 0.801 and b-IoU of 0.494 with PointNu. However, their performance across other evaluation metrics was inconsistent.
Using both augmented neutrophils and macrophages, NucleiMix, by and large, outperformed the baseline and the other three augmentation methods for both backbone models, providing substantial improvements across most evaluation metrics with the highest scores for DICE, AJI, DQ, and PQ for both model and b-IoU for SONNET. While CopyPaste obtained the best SQ scores for both models and best b-IoU for PointNu, its performance was inferior to NucleiMix for all other metrics. The other two augmentation methods not only performed worse than NucleiMix but also fell below the baseline on certain metrics.

The results of nuclei classification on MoNuSAC with augmented macrophages, neutrophils, and both are available in Table \ref{tab:cls}.
Provided with augmented macrophages, NucleiMix was superior to the baseline as well as the three augmentation methods, achieving the highest scores across all evaluation metrics. Notably, concerning $F_1$ for macrophages, NucleiMix provided substantial performance gains, with increases of 4.11\% and 8.78\% when using SONNET and PointNu, respectively. While the other three augmentation methods also improved $F^M_c$ compared to the baseline, their improvements were smaller than those obtained by NucleiMix, and, in some cases, they obtained worse performance than the baseline on certain evaluation metrics.
Using augmented neutrophils, the performance of NucleiMix was susceptible to the choice of the backbone model. As paired with PointNu, NucleiMix exhibited the highest scores across all evaluation metrics; however, with SONNET, it only excelled in $F_1$ for macrophages, and no other method demonstrated consistent superiority with SONNET. For example, the baseline obtained the highest scores for the overall $F_d$ (0.830) and $F^N_c$ (0.514). GradMix showed the best $F_1$ scores for $F^E_c$ (0.729) and $F^L_c$ (0.795). CutMix recorded the highest $Acc$ (0.947). It was noteworthy that no augmentation method surpassed the baseline $F^N_c$.
Augmenting both macrophages and neutrophils, NucleiMix was shown to be the top-performing method compared to the baseline and the other three augmentation methods. With SONNET, NucleiMix achieved the top scores across all evaluation metrics. Similarly, with PointNu, it obtained the best scores in three evaluation metrics ($F_d$, $F^M_c$, and $F^N_c$) and the second-best scores for $F^L_c$ and $F^E_c$ It is noteworthy that NucleiMix provided substantial improvements over the baseline for macrophages and neutrophils, increasing $F^M_c$ and $F^N_c$ by 7.24\% and 4.67\% with SONNET, and by 6.59\% and 4.33\% with PointNu, respectively.

\begin{table*}[htbp]
\caption{Effect of Number of Augmented Nuclei for Nuclei Segmentation} 
\label{tab:eff_seg}
\begin{center}    
\resizebox{\linewidth}{!}{
\begin{tabular}{cccccccccccc} 
\hline
\rule[-1ex]{0pt}{3.5ex} Model & Augmentation  & DICE  & \(\uparrow\) (\%)& AJI & \(\uparrow\) (\%)  & DQ & \(\uparrow\) (\%)  & SQ  & \(\uparrow\) (\%) & PQ & \(\uparrow\) (\%)   \\ 
\hline
\rule[-1ex]{0pt}{3.5ex} \multirow{6}{6em}{SONNET\\(CoNSeP: Miscellaneous)} & Baseline  & 0.830$\pm$0.001 && 0.479$\pm$0.009 && 0.632$\pm$0.006 && 0.758$\pm$0.002 && 0.480$\pm$0.006 & \\
\cline{2-12}
\rule[-1ex]{0pt}{3.5ex}  & k=200 & 0.830$\pm$0.001 & 0.00 & 0.482$\pm$0.003 &0.63& 0.641$\pm$0.002 &1.42& 0.763$\pm$0.004 &0.66& 0.480$\pm$0.006 &0.00  \\

\rule[-1ex]{0pt}{3.5ex}  & k=400 & $\mathbf{0.832\pm0.001}$ &$\mathbf{0.24}$& 0.495$\pm$0.005 &3.34& 0.647$\pm$0.005 &2.37& $\mathbf{0.763\pm0.002}$ &$\mathbf{0.66}$& 0.495$\pm$0.005 &3.13   \\

\rule[-1ex]{0pt}{3.5ex}  & k=600 & $\mathbf{0.832\pm0.001}$ &$\mathbf{0.24}$& $\mathbf{0.498\pm0.006}$ &$\mathbf{3.97}$& $\mathbf{0.651\pm0.005}$ &$\mathbf{3.01}$& $\mathbf{0.763\pm0.002}$ &$\mathbf{0.66}$& $\mathbf{0.498\pm0.006}$ &$\mathbf{3.75}$ \\

\rule[-1ex]{0pt}{3.5ex}  & k=800 & 0.829$\pm$0.001 &-0.12& 0.486$\pm$0.005 &1.46& 0.640$\pm$0.000 &1.27& 0.761$\pm$0.001 &0.40& 0.488$\pm$0.001 &1.67 \\
\hline

\rule[-1ex]{0pt}{3.5ex} \multirow{6}{6em}{PointNu\\(CoNSeP: Miscellaneous)} & Baseline  & 0.743$\pm$0.005 && 0.491$\pm$0.006 && 0.665$\pm$0.001 && 0.757$\pm$0.002 && 0.504$\pm$0.001 & \\
\cline{2-12}
\rule[-1ex]{0pt}{3.5ex}  & k=200 & 0.772$\pm$0.009 &3.90& 0.517$\pm$0.006 &5.30& 0.683$\pm$0.007 &2.71& 0.758$\pm$0.003 &0.13& 0.520$\pm$0.006 & 3.17  \\

\rule[-1ex]{0pt}{3.5ex}  & k=400 & $\mathbf{0.785\pm0.005}$ &$\mathbf{5.65}$& 0.523$\pm$0.006 &6.52& $\mathbf{0.689\pm0.002}$ &$\mathbf{3.61}$& $\mathbf{0.760\pm0.002}$ &$\mathbf{0.40}$& $\mathbf{0.525\pm0.002}$ &$\mathbf{4.17}$ \\

\rule[-1ex]{0pt}{3.5ex}  & k=600 & 0.784$\pm$0.004 &5.52& $\mathbf{0.525\pm0.001}$ &$\mathbf{6.92}$& 0.685$\pm$0.004 &3.01& 0.757$\pm$0.002 &0.00& 0.520$\pm$0.005 &3.17 \\

\rule[-1ex]{0pt}{3.5ex}  & k=800 & 0.781$\pm$0.005 &5.11& 0.522$\pm$0.004 &6.31& 0.680$\pm$0.006 &2.26& 0.755$\pm$0.005 &-0.26& 0.514$\pm$0.008 &1.98  \\
\hline

\rule[-1ex]{0pt}{3.5ex} \multirow{6}{6em}{SONNET\\(GLySAC: Miscellaneous)} & Baseline  & 0.822$\pm$0.002 && 0.627$\pm$0.004 && 0.786$\pm$0.005 && 0.782$\pm$0.002 && 0.617$\pm$0.004 & \\
\cline{2-12}
\rule[-1ex]{0pt}{3.5ex}  & k=200 & 0.823$\pm$0.001 &0.12& 0.632$\pm$0.002 &0.80& 0.785$\pm$0.002 &-0.13& $\mathbf{0.785\pm0.001}$ &$\mathbf{0.38}$& 0.619$\pm$0.002 &0.32  \\

\rule[-1ex]{0pt}{3.5ex}  & k=400 & $\mathbf{0.824\pm0.001}$ &$\mathbf{0.24}$& $\mathbf{0.634\pm0.004}$ &$\mathbf{1.12}$& $\mathbf{0.788\pm0.005}$&$\mathbf{0.25}$& $\mathbf{0.785\pm0.001}$ &$\mathbf{0.38}$& $\mathbf{0.621\pm0.004}$ &$\mathbf{0.65}$   \\

\rule[-1ex]{0pt}{3.5ex}  & k=600 & 0.822$\pm$0.000 &0.00& 0.631$\pm$0.001 &0.64& 0.786$\pm$0.004 &0.00& 0.784$\pm$0.002 &0.26& 0.619$\pm$0.002 &0.32 \\

\rule[-1ex]{0pt}{3.5ex}  & k=800 & 0.823$\pm$0.001 &0.12& 0.632$\pm$0.002 &0.80& 0.787$\pm$0.004 &0.13& $\mathbf{0.785\pm0.001}$ &$\mathbf{0.38}$& 0.620$\pm$0.004 &0.49  \\
\hline

\rule[-1ex]{0pt}{3.5ex} \multirow{6}{6em}{PointNu\\(GLySAC: Miscellaneous)} & Baseline  & 0.808$\pm$0.001 && 0.629$\pm$0.001 && 0.800$\pm$0.001 && $\mathbf{0.786\pm0.000}$ && 0.630$\pm$0.001 & \\
\cline{2-12}
\rule[-1ex]{0pt}{3.5ex}  & k=200 & 0.805$\pm$0.003 &-0.37& 0.624$\pm$0.002 &-0.79& 0.799$\pm$0.002 &-0.13& 0.782$\pm$0.002 &-0.51& 0.626$\pm$0.003 &-0.63  \\

\rule[-1ex]{0pt}{3.5ex}  & k=400 & 0.806$\pm$0.003 &-0.25& 0.623$\pm$0.005 &-0.95& 0.797$\pm$0.006 &-0.38& 0.781$\pm$0.001 &-0.64& 0.624$\pm$0.005 &-0.95   \\

\rule[-1ex]{0pt}{3.5ex}  & k=600 & 0.808$\pm$0.001 &0.00& 0.629$\pm$0.001 &0.00& 0.802$\pm$0.001 &0.25& 0.786$\pm$0.001 &-0.13& $\mathbf{0.631\pm0.000}$ &$\mathbf{0.16}$\\

\rule[-1ex]{0pt}{3.5ex}  & k=800 & $\mathbf{0.809\pm0.004}$ &$\mathbf{0.12}$& $\mathbf{0.630\pm0.003}$ &$\mathbf{0.16}$& $\mathbf{0.803\pm0.002}$ &$\mathbf{0.38}$& 0.782$\pm$0.002 &-0.51& 0.629$\pm$0.002 &-0.16  \\
\hline

\rule[-1ex]{0pt}{3.5ex} \multirow{6}{6em}{SONNET\\(MoNuSAC: Macrophages)} & Baseline  & 0.774$\pm$0.006 && 0.624$\pm$0.006 && 0.788$\pm$0.007 && 0.797$\pm$0.003 && 0.643$\pm$0.004 & \\
\cline{2-12}
\rule[-1ex]{0pt}{3.5ex}  & k=200 & 0.779$\pm$0.003 &0.65& 0.631$\pm$0.003 &1.12& $\mathbf{0.795\pm0.003}$ &$\mathbf{0.89}$& 0.811$\pm$0.010 &1.76& $\mathbf{0.650\pm0.003}$ &$\mathbf{1.09}$  \\

\rule[-1ex]{0pt}{3.5ex}  & k=400 & 0.780$\pm$0.002 &0.78& $\mathbf{0.633\pm0.003}$ &$\mathbf{1.44}$& 0.792$\pm$0.007 &0.51& 0.804$\pm$0.006 &0.88& 0.648$\pm$0.004 &0.78   \\

\rule[-1ex]{0pt}{3.5ex}  & k=600 & $\mathbf{0.782\pm0.001}$ &$\mathbf{1.03}$& 0.630$\pm$0.004 &0.96& 0.795$\pm$0.005 &0.89& $\mathbf{0.816\pm0.003}$ &$\mathbf{2.38}$& 0.650$\pm$0.004 &1.09 \\

\rule[-1ex]{0pt}{3.5ex}  & k=800 & 0.776$\pm$0.002 &0.26& 0.625$\pm$0.005 &0.16& 0.787$\pm$0.003 &-0.13& 0.802$\pm$0.006 &0.63& 0.643$\pm$0.000 &0.00  \\
\hline

\rule[-1ex]{0pt}{3.5ex} \multirow{6}{6em}{PointNu\\(MoNuSAC: Macrophages)} & Baseline  & 0.742$\pm$0.010 && 0.600$\pm$0.009 && 0.777$\pm$0.009 && 0.797$\pm$0.005 && 0.635$\pm$0.008 & \\
\cline{2-12}
\rule[-1ex]{0pt}{3.5ex}  & k=200 & 0.752$\pm$0.004 &1.35& 0.609$\pm$0.006 &1.50& 0.789$\pm$0.010 &1.54& 0.797$\pm$0.003 &0.00& 0.644$\pm$0.006 &1.42  \\

\rule[-1ex]{0pt}{3.5ex}  & k=400 & $\mathbf{0.766\pm0.004}$ &$\mathbf{3.23}$& $\mathbf{0.624\pm0.005}$ &$\mathbf{4.00}$& $\mathbf{0.799\pm0.002}$ &$\mathbf{2.83}$& $\mathbf{0.800\pm0.000}$ &$\mathbf{0.38}$& $\mathbf{0.655\pm0.002}$ &$\mathbf{3.15}$   \\

\rule[-1ex]{0pt}{3.5ex}  & k=600 & 0.756$\pm$0.004 &1.89& 0.614$\pm$0.004 &2.33& 0.790$\pm$0.004 &1.67& 0.799$\pm$0.004 &0.25& 0.647$\pm$0.002 &1.89 \\

\rule[-1ex]{0pt}{3.5ex}  & k=800 & 0.754$\pm$0.007 &1.62& 0.611$\pm$0.007 &1.83& 0.782$\pm$0.007 &0.64& $\mathbf{0.800\pm0.000}$ &$\mathbf{0.38}$& 0.641$\pm$0.006 &0.94  \\
\hline

\rule[-1ex]{0pt}{3.5ex} \multirow{6}{6em}{SONNET\\(MoNuSAC: Neutrophils)} & Baseline  & 0.774$\pm$0.006 && 0.624$\pm$0.006 && 0.788$\pm$0.007 && 0.797$\pm$0.003 && 0.643$\pm$0.004 & \\
\cline{2-12}
\rule[-1ex]{0pt}{3.5ex}  & k=200 & $\mathbf{0.781\pm0.003}$ &$\mathbf{0.90}$& $\mathbf{0.634\pm0.001}$ &$\mathbf{1.60}$& 0.794$\pm$0.001 &0.76& 0.814$\pm$0.009 &2.13& $\mathbf{0.651\pm0.001}$ &$\mathbf{1.24}$  \\

\rule[-1ex]{0pt}{3.5ex}  & k=400 & 0.778$\pm$0.007 &0.52& 0.630$\pm$0.005 &0.96& 0.791$\pm$0.002 &0.38& $\mathbf{0.820\pm0.001}$ &$\mathbf{2.89}$& 0.649$\pm$0.001 &0.93   \\

\rule[-1ex]{0pt}{3.5ex}  & k=600 & 0.778$\pm$0.002 &0.52& 0.627$\pm$0.001 &0.48& 0.792$\pm$0.001 &0.51& 0.819$\pm$0.001 &2.76& 0.650$\pm$0.001 &1.09 \\

\rule[-1ex]{0pt}{3.5ex}  & k=800 & 0.779$\pm$0.005 &0.65& 0.628$\pm$0.001 &0.64& $\mathbf{0.795\pm0.005}$ &$\mathbf{0.89}$& 0.807$\pm$0.009 &1.25& 0.647$\pm$0.002 &0.62  \\
\hline

\rule[-1ex]{0pt}{3.5ex} \multirow{5}{6em}{PointNu\\(MoNuSAC: Neutrophils)} & Baseline  & 0.742$\pm$0.010 && 0.600$\pm$0.009 && 0.777$\pm$0.009 && 0.797$\pm$0.005 && 0.635$\pm$0.008 & \\
\cline{2-12}
\rule[-1ex]{0pt}{3.5ex}  & k=200 & 0.750$\pm$0.005 &1.08& 0.609$\pm$0.006 &1.50& 0.783$\pm$0.006 &0.77& $\mathbf{0.800\pm0.002}$ &$\mathbf{0.38}$& 0.642$\pm$0.005 &1.10  \\

\rule[-1ex]{0pt}{3.5ex}  & k=400 & 0.761$\pm$0.006 &2.56& 0.619$\pm$0.007 &3.17& 0.790$\pm$0.004 &1.67& 0.798$\pm$0.002 &0.13& 0.646$\pm$0.002 &1.73   \\

\rule[-1ex]{0pt}{3.5ex}  & k=600 & $\mathbf{0.764\pm0.005}$ &$\mathbf{2.96}$& $\mathbf{0.624\pm0.007}$ &$\mathbf{4.00}$& $\mathbf{0.797\pm0.003}$ &$\mathbf{2.57}$& 0.799$\pm$0.002 &0.25& $\mathbf{0.653\pm0.004}$ &$\mathbf{2.83}$ \\

\rule[-1ex]{0pt}{3.5ex}  & k=800 & 0.760$\pm$0.004 &2.43& 0.618$\pm$0.007 &3.00& 0.794$\pm$0.005 &2.19& 0.798$\pm$0.002 &0.13& 0.649$\pm$0.006 &2.20  \\
\hline

\rule[-1ex]{0pt}{3.5ex} \multirow{6}{6em}{SONNET\\(MoNuSAC: Macrophages and Neutrophils)} & Baseline  & 0.774$\pm$0.006 && 0.624$\pm$0.006 && 0.788$\pm$0.007 && 0.797$\pm$0.003 && 0.643$\pm$0.004 & \\
\cline{2-12}
\rule[-1ex]{0pt}{3.5ex}  & k=200 & 0.775$\pm$0.003 &-0.13& 0.628$\pm$0.002 &0.64& $\mathbf{0.795\pm0.0032}$ &$\mathbf{0.89}$& $\mathbf{0.812\pm0.011}$ &$\mathbf{1.88}$& $\mathbf{0.652\pm0.004}$ &$\mathbf{1.40}$  \\

\rule[-1ex]{0pt}{3.5ex}  & k=400 & 0.777$\pm$0.003 &0.39& 0.628$\pm$0.007 &0.64& 0.788$\pm$0.003 &0.00& 0.803$\pm$0.004 &0.75& 0.644$\pm$0.003 &0.16   \\

\rule[-1ex]{0pt}{3.5ex}  & k=600 & $\mathbf{0.781\pm0.005}$ &$\mathbf{0.90}$& $\mathbf{0.632\pm0.007}$ &$\mathbf{1.28}$& 0.794$\pm$0.008 &0.76& 0.809$\pm$0.006 &1.51& 0.649$\pm$0.004 &0.93 \\

\rule[-1ex]{0pt}{3.5ex}  & k=800 & 0.775$\pm$0.003 &0.13& 0.626$\pm$0.004 &0.32& 0.794$\pm$0.004 &0.76& 0.811$\pm$0.011 &1.76& 0.649$\pm$0.003 &0.93  \\
\hline

\rule[-1ex]{0pt}{3.5ex} \multirow{6}{6em}{PointNu\\(MoNuSAC: Macrophages and Neutrophils)} & Baseline  & 0.742$\pm$0.010 && 0.600$\pm$0.009 && 0.777$\pm$0.009 && 0.797$\pm$0.005 && 0.635$\pm$0.008 & \\
\cline{2-12}
\rule[-1ex]{0pt}{3.5ex}  & k=200 & $\mathbf{0.758\pm0.002}$ &$\mathbf{2.16}$& 0.614$\pm$0.000 &2.33& 0.787$\pm$0.003 &1.29& $\mathbf{0.800\pm0.002}$ &$\mathbf{0.38}$& 0.646$\pm$0.001 &1.73  \\

\rule[-1ex]{0pt}{3.5ex}  & k=400 & 0.758$\pm$0.010 &2.16& 0.612$\pm$0.013 &2.00& 0.788$\pm$0.013 &1.42& 0.798$\pm$0.003 &0.13& 0.645$\pm$0.012 &1.57   \\

\rule[-1ex]{0pt}{3.5ex}  & k=600 & 0.757$\pm$0.000 &2.02& $\mathbf{0.616\pm0.002}$ &$\mathbf{2.67}$& $\mathbf{0.795\pm0.002}$ &$\mathbf{2.32}$& 0.796$\pm$0.000 &-0.13& $\mathbf{0.649\pm0.002}$ &$\mathbf{2.20}$ \\

\rule[-1ex]{0pt}{3.5ex}  & k=800 & 0.749$\pm$0.004 &0.94& 0.603$\pm$0.005 &0.50& 0.787$\pm$0.002 &1.29& 0.793$\pm$0.005 &-0.50& 0.640$\pm$0.003 &0.79  \\
\hline
\end{tabular}
}
\end{center}
\end{table*}

\subsection{Ablation Study: Effect of Number of Augmented Nuclei on Nuclei Segmentation}
To investigate the effect of the number of augmented rare-type nuclei, we repeated the nuclei segmentation experiments with varying k for each dataset. The results of nuclei segmentation on CoNSeP with augmented miscellaneous nuclei, GLySAC with augmented miscellaneous nuclei, and MoNuSAC with augmented macrophages, neutrophils, and both macrophages and neutrophils are available in Table \ref{tab:eff_seg}.

On CoNSeP, the addition of synthesized miscellaneous generally provided a performance gain over the baseline across all evaluation metrics, regardless of the number of augmented miscellaneous nuclei. With SONNET, the best performance was achieved when k=600, with the improvements ranging from 0.24\% to 3.97\%, while, with PointNu, it was k=400, which yielded the highest scores for DICE (0.785), DQ (0.689), SQ (0.760), and PQ (0.525).

On GLySAC, the impact of the augmented miscellaneous nuclei varied depending on the backbone models. With SONNET, NucleiMix consistently outperformed the baseline across all evaluation metrics, irrespective of the number of augmented nuclei, except for DQ at k=200. The optimal performance was observed at k=400. In contrast, with PointNu, the performance decreased at both k=200 and k=400. While the highest improvements in DICE (0.809), AJI (0.630), and DQ (0.803) were acquired at k=800, the highest PQ (0.631) was achieved at k=600.

On MoNuSAC, the addition of synthesized nuclei clearly enhanced the quality of nuclei segmentation. 
By augmenting neutrophils alone, both models demonstrated superior performance compared to the baseline across all evaluation metrics. With SONNET, the highest scores for DICE, AJI, and PQ were achieved at k=200, with improvements ranging from 0.90\% to 1.60\%. With PointNu, the best performance across all evaluation metrics, except SQ, was obtained at k=600, showing improvements between 2.57\% and 4.00\%.
Similarly, with respect to macrophages, NucleiMix consistently improved the performance of nuclei segmentation with a varying number of augmented macrophages, regardless of backbone models, except for DQ with SONNET when k=800. Using SONNET, the highest DQ and PQ were achieved at k=200 and the highest DICE and SQ were obtained at k=600. For PointNu, the best scores were attained at k=400 across all evaluation metrics, which improves upon the performance between 0.38\% and 4.00\%. 
As for augmenting both macrophages and neutrophils, the strength of NucleiMix was also evident. Varying k, the segmentation performance was consistently improved compared to the baseline across all evaluation metrics, except for DICE with SONNET at k=200 and for SQ with PointNu at k=600 and k=800. With SONNET, the highest improvements were acquired for DQ, SQ, and PQ with gains of 0.89\%, 1.88\%, and 1.40\%, respectively, at k=200, and at k=600 for DICE (0.90\%) and AJI (1.28\%). Using PointNu, when k=600, the greatest improvements of 2.67\%, 2.32\%, and 2.20\% were obtained for AJI, DQ, and PQ, respectively, while DICE and SQ enhanced by 2.16\% and 0.38\% at k=200. 

\begin{table*}[htbp]
\caption{Effect of Number of Augmented Nuclei for Nuclei Classification} 
\label{tab:eff_cls}
\begin{center}    
\resizebox{\linewidth}{!}{
\begin{tabular}{cccccccccccccc} 
\hline

\rule[-1ex]{0pt}{3.5ex} Model & Augmentation  & $F_d$  & \(\uparrow\) (\%)& $Acc$ & \(\uparrow\) (\%)  & $F^M_c$ & \(\uparrow\) (\%)  & $F^I_c$  & \(\uparrow\) (\%) & $F^E_c$ & \(\uparrow\) (\%)   & $F^S_c$  & \(\uparrow\) (\%)  \\ 
\hline
\rule[-1ex]{0pt}{3.5ex} \multirow{6}{6em}{SONNET\\(CoNSeP: Miscellaneous)} & Baseline  & 0.741$\pm$0.003 && 0.852$\pm$0.003 && 0.352$\pm$0.051 && 0.595$\pm$0.006 && 0.607$\pm$0.010 && 0.555$\pm$0.008 &\\
\cline{2-14}
\rule[-1ex]{0pt}{3.5ex}  & k=200 & 0.733$\pm$0.002 &-1.08& 0.854$\pm$0.002 &0.23& 0.378$\pm$0.014 &7.39& $\mathbf{0.603\pm0.006}$ &$\mathbf{1.34}$& 0.601$\pm$0.006 &-0.99& 0.547$\pm$0.005 & -1.44   \\

\rule[-1ex]{0pt}{3.5ex}  & k=400 & $\mathbf{0.745\pm0.005}$ &$\mathbf{0.54}$& 0.854$\pm$0.005 &0.23& $\mathbf{0.422\pm0.018}$&$\mathbf{19.89}$& 0.602$\pm$0.018 &1.18& $\mathbf{0.618\pm0.012}$ &$\mathbf{1.81}$& 0.547$\pm$0.007 &-1.44   \\

\rule[-1ex]{0pt}{3.5ex}  & k=600 & 0.742$\pm$0.002 &0.13& $\mathbf{0.855\pm0.002}$ &$\mathbf{0.35}$& 0.410$\pm$0.041 &16.48& 0.598$\pm$0.016 &0.50& 0.608$\pm$0.004 &0.16& $\mathbf{0.555\pm0.005}$  &$\mathbf{0.00}$ \\

\rule[-1ex]{0pt}{3.5ex}  & k=800 & 0.725$\pm$0.003 &-2.16& 0.851$\pm$0.003 &-0.12& 0.383$\pm$0.025 &8.81& 0.598$\pm$0.014 &0.50& 0.580$\pm$0.007 &-4.45& 0.544$\pm$0.006  &-1.98 \\
\hline

\rule[-1ex]{0pt}{3.5ex} \multirow{6}{6em}{PointNu\\(CoNSeP: Miscellaneous)} & Baseline  & 0.724$\pm$0.003 && 0.888$\pm$0.005 && 0.544$\pm$0.021 && 0.661$\pm$0.020 && 0.620$\pm$0.014 && 0.542$\pm$0.014 &\\
\cline{2-14}
\rule[-1ex]{0pt}{3.5ex}  & k=200 & 0.741$\pm$0.007 &2.35& 0.892$\pm$0.001 &0.45& 0.576$\pm$0.008 &5.88& 0.670$\pm$0.006 &1.36& 0.646$\pm$0.009 & 4.19& 0.554$\pm$0.003  &2.21 \\

\rule[-1ex]{0pt}{3.5ex}  & k=400 & $\mathbf{0.751\pm0.004}$ &$\mathbf{3.73}$& 0.892$\pm$0.003 &0.45& 0.585$\pm$0.012 &7.54& 0.669$\pm$0.010 &1.21& $\mathbf{0.657\pm0.007}$ &$\mathbf{5.97}$& 0.561$\pm$0.007 &3.51\\

\rule[-1ex]{0pt}{3.5ex}  & k=600 & 0.744$\pm$0.004 &2.76& $\mathbf{0.897\pm0.007}$ &$\mathbf{1.01}$& $\mathbf{0.600\pm0.005}$ &$\mathbf{10.29}$& $\mathbf{0.680\pm0.006}$ &$\mathbf{2.87}$& 0.651$\pm$0.009 &5.00& 0.558$\pm$0.019 &2.95  \\

\rule[-1ex]{0pt}{3.5ex}  & k=800 & 0.744$\pm$0.005 &2.76& 0.892$\pm$0.002 &0.45& 0.578$\pm$0.004 &6.25& 0.664$\pm$0.011 &0.45& 0.646$\pm$0.011 &4.19& $\mathbf{0.562\pm0.002}$ &$\mathbf{3.69}$  \\
\hline

\rule[-1ex]{0pt}{3.5ex} Model & Augmentation  & $F_d$  & \(\uparrow\) (\%)& $Acc$ & \(\uparrow\)(\%)& $F^M_c$ & \(\uparrow\) (\%)& $F^L_c$    & \(\uparrow\) (\%)& $F^E_c$  & \(\uparrow\) (\%)\\ 
\hline
\rule[-1ex]{0pt}{3.5ex} \multirow{6}{6em}{SONNET\\(GLySAC: Miscellaneous)} & Baseline  & 0.838$\pm$0.003 && 0.698$\pm$0.004 && 0.287$\pm$0.05 && 0.520$\pm$0.007 && 0.540$\pm$0.005  &\\
\cline{2-12}
\rule[-1ex]{0pt}{3.5ex}  & k=200 & 0.839$\pm$0.005 &0.12& $\mathbf{0.707\pm0.002}$ &$\mathbf{1.29}$& 0.305$\pm$0.005 &6.27& 0.529$\pm$0.002 &1.73& 0.545$\pm$0.007 &0.93  \\
\rule[-1ex]{0pt}{3.5ex}  & k=400 & $\mathbf{0.844\pm0.004}$ &$\mathbf{0.72}$& 0.704$\pm$0.002 &0.86& 0.309$\pm$0.008 &7.67& 0.523$\pm$0.011 &0.58& $\mathbf{0.549\pm0.011}$  &$\mathbf{1.67}$\\
\rule[-1ex]{0pt}{3.5ex}  & k=600 & 0.843$\pm$0.004 &0.60& 0.706$\pm$0.005 &1.15& $\mathbf{0.310\pm0.011}$ &$\mathbf{8.01}$& $\mathbf{0.531\pm0.013}$ &$\mathbf{2.12}$& 0.545$\pm$0.003 &0.93  \\
\rule[-1ex]{0pt}{3.5ex}  & k=800 & 0.840$\pm$0.005 &0.24& 0.702$\pm$0.003 &0.57& 0.301$\pm$0.009 &4.88& 0.520$\pm$0.007 &0.00& 0.545$\pm$0.002 &0.93  \\
\hline
\rule[-1ex]{0pt}{3.5ex} \multirow{6}{6em}{PointNu\\(GLySAC: Miscellaneous)} & Baseline  & 0.824$\pm$0.002 && 0.715$\pm$0.010 && 0.294$\pm$0.010 && 0.518$\pm$0.019 && 0.553$\pm$0.004 &\\
\cline{2-12}
\rule[-1ex]{0pt}{3.5ex}  & k=200 & 0.833$\pm$0.005 &1.09& 0.724$\pm$0.010 &1.26& 0.307$\pm$0.011 &4.42& 0.536$\pm$0.021 &3.47& 0.566$\pm$0.007  &2.35 \\
\rule[-1ex]{0pt}{3.5ex}  & k=400 & 0.832$\pm$0.004 &0.97& 0.720$\pm$0.006 &0.70& 0.314$\pm$0.006 &6.80& 0.529$\pm$0.008 &2.12& 0.561$\pm$0.008  &1.45 \\
\rule[-1ex]{0pt}{3.5ex}  & k=600 & 0.832$\pm$0.005 &0.97& 0.720$\pm$0.006 &0.70& $\mathbf{0.318\pm0.004}$ &$\mathbf{8.16}$& 0.525$\pm$0.005 &1.35& 0.563$\pm$0.010 &1.81  \\
\rule[-1ex]{0pt}{3.5ex}  & k=800 & $\mathbf{0.835\pm0.005}$ &$\mathbf{1.33}$& $\mathbf{0.725\pm0.004}$ &$\mathbf{1.40}$& 0.309$\pm$0.005 &5.10& $\mathbf{0.536\pm0.007}$ &$\mathbf{3.47}$& $\mathbf{0.571\pm0.005}$  &$\mathbf{3.25}$ \\
\hline

\rule[-1ex]{0pt}{3.5ex} Model & Augmentation  & $F_d$  &\(\uparrow\) (\%) & $Acc$  &\(\uparrow\) (\%) & $F^E_c$ &\(\uparrow\) (\%) & $F^L_c$  &\(\uparrow\) (\%) & $F^M_c$  &\(\uparrow\) (\%) & $F^N_c$  &\(\uparrow\) (\%) \\ 
\hline
\rule[-1ex]{0pt}{3.5ex} \multirow{6}{6em}{SONNET\\(MoNuSAC: Macrophages)} & Baseline  & 0.830$\pm$0.001 && 0.939$\pm$0.004 && 0.709$\pm$0.005 && 0.786$\pm$0.008 && 0.511$\pm$0.017 && 0.514$\pm$0.028 &\\
\cline{2-14}
\rule[-1ex]{0pt}{3.5ex}  & k=200 & 0.832$\pm$0.008 &0.24& 0.940$\pm$0.008 &0.11& 0.719$\pm$0.006 &1.41& 0.787$\pm$0.010 &0.13& 0.536$\pm$0.013 &4.89& 0.526$\pm$0.033 &2.23  \\
\rule[-1ex]{0pt}{3.5ex}  & k=400 & $\mathbf{0.836\pm0.004}$ &$\mathbf{0.72}$& 0.948$\pm$0.003 &0.96& 0.736$\pm$0.004 &3.81& 0.798$\pm$0.007 &1.53& $\mathbf{0.565\pm0.011}$ &$\mathbf{10.57}$& $\mathbf{0.544\pm0.029}$  &$\mathbf{5.84}$ \\
\rule[-1ex]{0pt}{3.5ex}  & k=600 & 0.835$\pm$0.002 &0.60& $\mathbf{0.952\pm0.007}$ &$\mathbf{1.38}$& $\mathbf{0.746\pm0.010}$ &$\mathbf{5.22}$& $\mathbf{0.801\pm0.006}$ &$\mathbf{1.92}$& 0.532$\pm$0.009 &4.11& 0.535$\pm$0.019  &4.09 \\
\rule[-1ex]{0pt}{3.5ex}  & k=800 & 0.829$\pm$0.006 &-0.12& 0.950$\pm$0.007 &1.17& 0.738$\pm$0.016 &4.09& 0.793$\pm$0.014 &0.89& 0.536$\pm$0.004 &4.89& 0.539$\pm$0.018  &4.86 \\
\hline
\rule[-1ex]{0pt}{3.5ex} \multirow{6}{6em}{PointNu\\(MoNuSAC: Macrophages)} & Baseline  & 0.835$\pm$0.001 && 0.975$\pm$0.001 && 0.786$\pm$0.004 && 0.828$\pm$0.007 && 0.501$\pm$0.018 && 0.508$\pm$0.036 \\
\cline{2-14}
\rule[-1ex]{0pt}{3.5ex}  & k=200 & 0.838$\pm$0.006 &0.36& $\mathbf{0.977\pm0.002}$ &$\mathbf{0.21}$& 0.793$\pm$0.002 &0.89& 0.832$\pm$0.009 &0.48& 0.535$\pm$0.018 &6.79& 0.511$\pm$0.014 &0.59  \\
\rule[-1ex]{0pt}{3.5ex}  & k=400 & 0.839$\pm$0.004 &0.48& 0.976$\pm$0.000 &0.10& 0.793$\pm$0.006 &0.89& 0.832$\pm$0.002 &0.48& 0.542$\pm$0.018 &8.18& 0.524$\pm$0.018  &3.15\\
\rule[-1ex]{0pt}{3.5ex}  & k=600 & 0.840$\pm$0.006 &0.60& 0.976$\pm$0.001 &0.10& $\mathbf{0.796\pm0.006}$ &$\mathbf{1.27}$& 0.832$\pm$0.003 &0.48& $\mathbf{0.545\pm0.003}$ &$\mathbf{8.78}$& $\mathbf{0.528\pm0.008 }$ &$\mathbf{3.94}$\\
\rule[-1ex]{0pt}{3.5ex}  & k=800 & $\mathbf{0.840\pm0.002}$ &$\mathbf{0.60}$& 0.976$\pm$0.003 &0.10& 0.794$\pm$0.006 &1.02& $\mathbf{0.833\pm0.001}$ &$\mathbf{0.60}$& 0.524$\pm$0.018 &4.59& 0.509$\pm$0.017 &0.20  \\
\hline

\rule[-1ex]{0pt}{3.5ex} \multirow{6}{6em}{SONNET\\(MoNuSAC: Neutrophils)} & Baseline  & 0.830$\pm$0.001 && 0.939$\pm$0.004 && 0.709$\pm$0.005 && 0.786$\pm$0.008 && 0.511$\pm$0.017 && 0.514$\pm$0.028 &\\
\cline{2-14}
\rule[-1ex]{0pt}{3.5ex}  & k=200 & 0.832$\pm$0.004 &0.20&  $\mathbf{0.957\pm0.001}$ &$\mathbf{1.92}$& $\mathbf{0.741\pm0.007}$ &$\mathbf{4.51}$& $\mathbf{0.811\pm0.006}$ &$\mathbf{3.18}$& $\mathbf{0.542\pm0.006}$ &$\mathbf{6.07}$& $\mathbf{0.532\pm0.005}$  &$\mathbf{3.50}$ \\
\rule[-1ex]{0pt}{3.5ex}  & k=400 & 0.832$\pm$0.003 &0.24& 0.950$\pm$0.004 &1.17& 0.740$\pm$0.013 &4.37& 0.799$\pm$0.018 &1.65& 0.529$\pm$0.012 &3.52& 0.517$\pm$0.007  &-0.78 \\
\rule[-1ex]{0pt}{3.5ex}  & k=600 & 0.827$\pm$0.003 &-0.36& 0.946$\pm$0.004 &0.75& 0.725$\pm$0.002 &2.26& 0.787$\pm$0.016 &0.13& 0.529$\pm$0.004 &3.52& 0.503$\pm$0.005  &-2.14 \\
\rule[-1ex]{0pt}{3.5ex}  & k=800 & $\mathbf{0.832\pm0.001}$ &$\mathbf{0.24}$& 0.943$\pm$0.005 &0.43& 0.721$\pm$0.006 &1.69& 0.794$\pm$0.014 &1.02& 0.514$\pm$0.007 &0.59& 0.525$\pm$0.006  &2.14 \\
\hline
\rule[-1ex]{0pt}{3.5ex} \multirow{6}{6em}{PointNu\\(MoNuSAC: Neutrophils)} & Baseline  & 0.835$\pm$0.001 && 0.975$\pm$0.001 && 0.786$\pm$0.004 && 0.828$\pm$0.007 && 0.501$\pm$0.018 && 0.508$\pm$0.036 &\\
\cline{2-14}
\rule[-1ex]{0pt}{3.5ex}  & k=200 & 0.838$\pm$0.005 &0.36& $\mathbf{0.978\pm0.000}$ &$\mathbf{0.31}$& $\mathbf{0.795\pm0.007}$ &$\mathbf{1.15}$& 0.832$\pm$0.009 &0.48& 0.537$\pm$0.027 &7.19& $\mathbf{0.550\pm0.019}$  &$\mathbf{8.27}$ \\
\rule[-1ex]{0pt}{3.5ex}  & k=400 & 0.839$\pm$0.005 &0.48& 0.976$\pm$0.000 &0.10& 0.788$\pm$0.004 &0.25& 0.832$\pm$0.007 &0.48& 0.552$\pm$0.035 &10.18& 0.528$\pm$0.003 &3.94  \\
\rule[-1ex]{0pt}{3.5ex}  & k=600 & 0.839$\pm$0.005 &0.48& 0.976$\pm$0.000 &0.10& 0.794$\pm$0.008 &1.02& 0.829$\pm$0.002 &0.12& 0.534$\pm$0.011 &6.59& 0.538$\pm$0.021 &5.91  \\
\rule[-1ex]{0pt}{3.5ex}  & k=800 & $\mathbf{0.840\pm0.004}$ &$\mathbf{0.60}$& 0.975$\pm$0.001 &0.00& 0.790$\pm$0.007 &0.51& $\mathbf{0.832\pm0.004}$ &0.48& $\mathbf{0.557\pm0.014}$ &$\mathbf{11.18}$& 0.515$\pm$0.008 &1.38  \\
\hline

\rule[-1ex]{0pt}{3.5ex} \multirow{6}{6em}{SONNET\\(MoNuSAC: Macrophages and Neutrophils)} & Baseline  & 0.830$\pm$0.001 && 0.939$\pm$0.004 && 0.709$\pm$0.005 && 0.786$\pm$0.008 && 0.511$\pm$0.017 && 0.514$\pm$0.028 &\\
\cline{2-14}
\rule[-1ex]{0pt}{3.5ex}  & k=200 & $\mathbf{0.834\pm0.003}$ &$\mathbf{0.48}$& 0.943$\pm$0.002 &0.43& 0.731$\pm$0.006 &3.10& 0.788$\pm$0.004 &0.25& 0.541$\pm$0.021 &5.87& 0.520$\pm$0.014 &1.17  \\
\rule[-1ex]{0pt}{3.5ex}  & k=400 & 0.829$\pm$0.003 &-0.12& 0.941$\pm$0.007 &0.21& 0.721$\pm$0.014 &1.69& 0.785$\pm$0.009 &-0.13& 0.524$\pm$0.010 &2.54& 0.518$\pm$0.020  &0.78 \\
\rule[-1ex]{0pt}{3.5ex}  & k=600 & 0.830$\pm$0.001 &0.00& $\mathbf{0.954\pm0.003}$ &$\mathbf{1.60}$& $\mathbf{0.740\pm0.018}$ &$\mathbf{4.37}$& $\mathbf{0.802\pm0.007}$ &$\mathbf{2.04}$& $\mathbf{0.548\pm0.014}$ &$\mathbf{7.24}$& $\mathbf{0.538\pm0.006}$  &$\mathbf{4.67}$ \\
\rule[-1ex]{0pt}{3.5ex}  & k=800 & 0.831$\pm$0.003 &0.12& 0.946$\pm$0.005 &0.75& 0.736$\pm$0.003 &3.81& 0.790$\pm$0.011 &0.51& 0.501$\pm$0.027 &-1.96& 0.518$\pm$0.016  &0.78 \\
\hline
\rule[-1ex]{0pt}{3.5ex} \multirow{6}{6em}{PointNu\\(MoNuSAC: Macrophages and Neutrophils)} & Baseline  & 0.835$\pm$0.001 && 0.975$\pm$0.001 && 0.786$\pm$0.004 && 0.828$\pm$0.007 && 0.501$\pm$0.018 && 0.508$\pm$0.036 \\
\cline{2-14}
\rule[-1ex]{0pt}{3.5ex}  & k=200 & 0.842$\pm$0.002 &0.84& $\mathbf{0.976\pm0.001}$ &$\mathbf{0.10}$& 0.797$\pm$0.001 &1.40& $\mathbf{0.835\pm0.002}$ &$\mathbf{0.85}$& 0.524$\pm$0.013 &4.59& 0.524$\pm$0.009 &3.15 \\
\rule[-1ex]{0pt}{3.5ex}  & k=400 & 0.839$\pm$0.004 &0.48& 0.975$\pm$0.001 &0.00& 0.794$\pm$0.008 &1.02& 0.830$\pm$0.002 &0.24& 0.534$\pm$0.013 &6.59& 0.526$\pm$0.015  &3.54 \\
\rule[-1ex]{0pt}{3.5ex}  & k=600 & $\mathbf{0.844\pm0.002}$ &$\mathbf{1.08}$& 0.974$\pm$0.003 &-0.10& $\mathbf{0.799\pm0.008}$ &$\mathbf{1.65}$& 0.829$\pm$0.007 &0.12& $\mathbf{0.534\pm0.018}$ &$\mathbf{6.59}$& $\mathbf{0.530\pm0.013}$  &$\mathbf{4.33}$ \\
\rule[-1ex]{0pt}{3.5ex}  & k=800 & 0.839$\pm$0.005 &0.48& 0.972$\pm$0.004 &-0.31& 0.789$\pm$0.012 &0.38& 0.824$\pm$0.011 &-0.48& 0.518$\pm$0.008 &3.39& 0.518$\pm$0.016  &1.97 \\
\hline

\end{tabular}
}
\end{center}
\end{table*}

\subsection{Ablation Study: Effect of Number of Augmented Nuclei on Nuclei Classification}
By augmenting rare-type nuclei, we also conducted the nuclei classification experiments with differing k for each dataset. Table \ref{tab:eff_cls} illustrates the results of nuclei classification on CoNSeP with augmented miscellaneous nuclei, GLySAC with augmented miscellaneous nuclei, and MoNuSAC with augmented neutrophils, macrophages, and both macrophages and neutrophils. The inclusion of synthesized rare-type nuclei generally improved classification performance regardless of the number of augmented nuclei and the type of nuclei. In particular, in most cases, the classification performance of rare-type nuclei showed substantial improvement. 

On CoNSeP, SONNET achieved the highest scores for $F_d$, $F^M_c$, and $F^E_c$ at k=400, $Acc$ and $F^S_c$ at k=600, and $F^I_c$ at k=200. Meanwhile, PointNu obtained the best scores for $Acc$, $F^M_c$, and $F^I_c$ at k=600, $F_d$ and $F^E_c$ at k=400, and $F^S_c$ at k=800. The greatest improvements for miscellaneous nuclei ($F^M_c$) were observed at k=400 with SONNET and at k=600 with PointNu, with improvements of 19.89\% and 10.29\%, respectively, compared to the baseline. 

On GLySAC, using SONNET, NucleiMix attained the most substantial improvements for $F_d$ and $F^E_c$ at k=400, $Acc$ at k=200, and $F^M_c$ and $F^L_c$ at k=600.
PointNu benefited the most from NucleiMix at k=800, achieving the highest $F_d$, $Acc$, $F^L_c$, and $F^E_c$. In regard to miscellaneous nuclei, the highest $F^M_c$ was obtained at k=600 for both models, with enhancements of $\ge$8.01\%.

On MoNuSAC, augmenting neutrophils alone led to the highest scores with SONNET at k=200 across all evaluation metrics except for $F_d$. As for PointNu, the best scores were split between k=200 ($Acc$, $F^E_c$, and $F^N_c$) and k=800 ($F_d$, $F^L_c$, and $F^M_c$). The neutrophil nuclei class showed the most improvement for SONNET and PointNu at k=200 by 3.50\% and 8.27\%, respectively. 
With the augmented macrophages only, k=400 and k=600 proved to be the most beneficial for SONNET, yielding the highest scores for metrics such as $F_d$, $F^M_c$, and $F^N_c$ at k=400 and $Acc$, $F^E_c$, and $F^L_c$ at k=600. For PointNu, k=600 was superior to others, resulting in the highest scores for $F^E_c$, $F^M_c$, and $F^N_c$. In the macrophages nuclei category, the best scores of 0.565 and 0.545 were achieved by SONNET at k=400 and PointNu at k=600, respectively.
Including both synthesized neutrophils and macrophages, both models, by and large, performed best at k=600; for instance, the most improvements were obtained for SONNET in $Acc$ (1.60\%), $F^E_c$ (4.37\%), $F^L_c$ (2.04\%), $F^M_c$ (7.24\%), and $F^N_c$ (4.67\%) and for PointNu in $F_d$ (1.08\%), $F^E_c$ (1.65\%), $F^M_c$ (6.59\%), and $F^N_c$ (4.33\%).

\begin{table*}[htbp]
\caption{Results of Class-wise Nuclei Segmentation (k=600)}
\label{tab:seg_class}
\begin{center}    
\resizebox{0.8\linewidth}{!}{
\begin{tabular}{ccccccccccccccc} 
\hline
\rule[-1ex]{0pt}{3.5ex} Model & Nuclei Type & Augmentation  & DICE  & \(\uparrow\) (\%)& AJI & \(\uparrow\) (\%)  & DQ & \(\uparrow\) (\%)  & SQ  & \(\uparrow\) (\%) & PQ & \(\uparrow\) (\%) & b-IoU  & \(\uparrow\) \\ 
\hline
\rule[-1ex]{0pt}{3.5ex} \multirow{8}{6em}{SONNET\\(CoNSeP: Miscellaneous)} 
&  \multirow{2}{6em}{miscellaneous}  & Baseline & 0.136±0.020 && 0.072±0.012 && 0.094±0.011 && 0.242±0.026 && 0.068±0.008 && 0.237±0.027 & \\
\rule[-1ex]{0pt}{3.5ex}  & & NucleiMix & 0.148±0.009 &9.09& 0.079±0.008 &8.76& 0.110±0.014 &16.96& 0.284±0.020 &17.33& 0.081±0.009 &19.61& 0.281±0.019 &18.26   \\
\cline{2-15}

\rule[-1ex]{0pt}{3.5ex}  & \multirow{2}{6em}{inflammatory} & Baseline & 0.532±0.025 && 0.356±0.021 && 0.479±0.010 && 0.772±0.029 && 0.386±0.005 && 0.763±0.028 &   \\
\rule[-1ex]{0pt}{3.5ex}  & & NucleiMix & 0.544±0.013 &2.38& 0.358±0.014 &0.66& 0.500±0.013 &4.39& 0.780±0.029 &1.08& 0.406±0.009 &5.09& 0.772±0.027 &1.09  \\
\cline{2-15}

\rule[-1ex]{0pt}{3.5ex}  & \multirow{2}{6em}{epithelial} & Baseline & 0.519±0.008 && 0.232±0.008 && 0.260±0.006 && 0.444±0.001 && 0.190±0.006 && 0.380±0.002 &   \\
\rule[-1ex]{0pt}{3.5ex}  & & NucleiMix & 0.523±0.007 &0.90& 0.243±0.006 &4.59& 0.255±0.010 &-1.92& 0.441±0.003 &-0.68& 0.185±0.008 &-2.28& 0.378±0.004 &-0.44  \\
\cline{2-15}

\rule[-1ex]{0pt}{3.5ex}  & \multirow{2}{6em}{spindle} & Baseline & 0.578±0.017 && 0.319±0.011 && 0.369±0.011 && 0.690±0.018 && 0.271±0.009 && 0.671±0.017 &   \\
\rule[-1ex]{0pt}{3.5ex}  & & NucleiMix & 0.589±0.015 &1.96& 0.322±0.011 &0.94& 0.375±0.015 &1.62& 0.707±0.003 &2.46& 0.274±0.012 &0.98& 0.685±0.003 &2.14 \\
\hline

\rule[-1ex]{0pt}{3.5ex} \multirow{8}{6em}{PointNu\\(CoNSeP: Miscellaneous)} 

& \multirow{2}{6em}{miscellaneous} & Baseline  & 0.129±0.004 && 0.071±0.002 && 0.118±0.006 && 0.323±0.007 && 0.088±0.005 && 0.320±0.006 & \\
\rule[-1ex]{0pt}{3.5ex}  & & NucleiMix & 0.156±0.013 &20.98& 0.084±0.009 &18.78& 0.128±0.013 &8.76& 0.307±0.010 &-4.86& 0.091±0.009 &3.41& 0.305±0.011 &-4.89  \\
\cline{2-15}

\rule[-1ex]{0pt}{3.5ex}  & \multirow{2}{6em}{inflammatory} & Baseline & 0.562±0.010 && 0.376±0.004 && 0.518±0.007 && 0.781±0.015 && 0.413±0.002 && 0.773±0.014 &    \\
\rule[-1ex]{0pt}{3.5ex}  & & NucleiMix & 0.594±0.002 &5.76& 0.394±0.004 &4.61& 0.524±0.006 &1.03& 0.790±0.007 &1.11& 0.419±0.004 &1.62& 0.782±0.006 &1.16 \\
\cline{2-15}

\rule[-1ex]{0pt}{3.5ex}  & \multirow{2}{6em}{epithelial} & Baseline & 0.462±0.008 && 0.261±0.006 && 0.307±0.018 && 0.446±0.002 && 0.223±0.014 && 0.385±0.001 &    \\
\rule[-1ex]{0pt}{3.5ex}  & & NucleiMix & 0.498±0.003 &7.64& 0.270±0.002 &3.45& 0.294±0.004 &-4.13& 0.443±0.001 &-0.75& 0.214±0.003 &-4.19& 0.381±0.001 &-1.12 \\
\cline{2-15}

\rule[-1ex]{0pt}{3.5ex}  & \multirow{2}{6em}{spindle} & Baseline & 0.520±0.009 && 0.299±0.008 && 0.381±0.016 && 0.665±0.023 && 0.274±0.011 && 0.648±0.023 &    \\
\rule[-1ex]{0pt}{3.5ex}  & & NucleiMix & 0.559±0.009 &7.50& 0.325±0.009 &8.58& 0.408±0.016 &7.09& 0.699±0.007 &5.01& 0.297±0.013 &8.27& 0.678±0.006 &4.58 \\
\hline
\rule[-1ex]{0pt}{3.5ex} \multirow{8}{6em}{SONNET\\(GLySAC: Miscellaneous} 

& \multirow{2}{6em}{miscellaneous} & Baseline  & 0.370±0.011 && 0.221±0.008 && 0.309±0.007 && 0.737±0.014 && 0.234±0.006 && 0.713±0.013 & \\
\rule[-1ex]{0pt}{3.5ex}  & & NucleiMix & 0.379±0.002 &2.43& 0.223±0.003 &1.21& 0.321±0.008 &3.89& 0.745±0.001 &1.13& 0.242±0.006 &3.57& 0.722±0.002 &1.26  \\
\cline{2-15}

\rule[-1ex]{0pt}{3.5ex}  & \multirow{2}{6em}{inflammation} & Baseline & 0.511±0.001 && 0.311±0.002 && 0.440±0.003 && 0.737±0.013 && 0.342±0.001 && 0.717±0.012 &\\
\rule[-1ex]{0pt}{3.5ex} & & NucleiMix & 0.513±0.010 &0.46& 0.315±0.009 &1.07& 0.445±0.004 &1.29& 0.745±0.002 &1.04& 0.347±0.005 &1.46& 0.725±0.002 &1.16  \\
\cline{2-15}

\rule[-1ex]{0pt}{3.5ex}  & \multirow{2}{6em}{epithelial} & Baseline & 0.581±0.007 && 0.351±0.006 && 0.403±0.007 && 0.693±0.007 && 0.313±0.007 && 0.631±0.007 & \\
\rule[-1ex]{0pt}{3.5ex}  & & NucleiMix & 0.584±0.006 &0.57& 0.355±0.006 &1.04& 0.403±0.007 &0.00& 0.697±0.002 &0.48& 0.314±0.007 &0.32& 0.635±0.003 &0.69   \\
\hline

\rule[-1ex]{0pt}{3.5ex} \multirow{8}{6em}{PointNu\\(GLySAC: Miscellaneous)} 

& \multirow{2}{6em}{miscellaneous} & Baseline  & 0.376±0.008 && 0.229±0.007 && 0.341±0.007 && 0.742±0.014 && 0.260±0.007 && 0.723±0.012 & \\
\rule[-1ex]{0pt}{3.5ex}  & & NucleiMix & 0.394±0.018 &4.61& 0.239±0.013 &4.66& 0.350±0.018 &2.64& 0.748±0.001 &0.85& 0.266±0.014 &2.05& 0.726±0.002 &0.41  \\
\cline{2-15}

\rule[-1ex]{0pt}{3.5ex}  & \multirow{2}{6em}{inflammation} & Baseline & 0.506±0.018 && 0.315±0.012 && 0.454±0.010 && 0.732±0.011 && 0.356±0.007 && 0.713±0.010 &\\
\rule[-1ex]{0pt}{3.5ex} & & NucleiMix & 0.507±0.016 &0.26& 0.312±0.009 &-0.85& 0.447±0.007 &-1.54& 0.733±0.013 &0.14& 0.347±0.005 &-2.43& 0.713±0.014 &0.05 \\
\cline{2-15}

\rule[-1ex]{0pt}{3.5ex}  & \multirow{2}{6em}{epithelial} & Baseline & 0.588±0.004 && 0.366±0.003 && 0.419±0.004 && 0.691±0.001 && 0.326±0.003 && 0.630±0.002 & \\
\rule[-1ex]{0pt}{3.5ex}  & & NucleiMix & 0.584±0.009 &-0.62& 0.373±0.002 &1.82& 0.437±0.008 &4.21& 0.700±0.006 &1.25& 0.344±0.009 &5.52& 0.638±0.006 &1.27 \\
\hline

\rule[-1ex]{0pt}{3.5ex} \multirow{8}{6em}{SONNET\\(MoNuSAC: Macrophages)} 

&  \multirow{2}{6em}{epithelial} & Baseline  & 0.242±0.004 && 0.175±0.004 && 0.209±0.007 && 0.282±0.019 && 0.169±0.006 && 0.203±0.011 &\\
\rule[-1ex]{0pt}{3.5ex}  & & NucleiMix & 0.246±0.008 &1.65& 0.179±0.006 &1.90& 0.212±0.007 &1.27& 0.272±0.016 &-3.43& 0.170±0.005 &0.59& 0.197±0.008 &-3.12  \\
\cline{2-15}

\rule[-1ex]{0pt}{3.5ex}  &  \multirow{2}{6em}{lymphocyte} & Baseline & 0.356±0.019 && 0.229±0.013 && 0.306±0.024 && 0.449±0.025 && 0.242±0.020 && 0.338±0.018 &\\
\rule[-1ex]{0pt}{3.5ex}  & & NucleiMix & 0.368±0.014 &3.37& 0.240±0.011 &4.96& 0.320±0.016 &4.58& 0.453±0.023 &0.97& 0.252±0.011 &3.99& 0.338±0.017 &0.10  \\
\cline{2-15}

\rule[-1ex]{0pt}{3.5ex}  &  \multirow{2}{6em}{macrophages} & Baseline & 0.167±0.008 && 0.130±0.007 && 0.150±0.008 && 0.194±0.019 && 0.128±0.006 && 0.090±0.008 &\\
\rule[-1ex]{0pt}{3.5ex}  & & NucleiMix & 0.173±0.003 &3.39& 0.136±0.006 &4.09& 0.156±0.005 &4.22& 0.217±0.005 &11.49& 0.132±0.005 &3.13& 0.094±0.002 &4.81 \\
\cline{2-15}

\rule[-1ex]{0pt}{3.5ex}  &  \multirow{2}{6em}{neutrophils} & Baseline & 0.205±0.022 && 0.165±0.020 && 0.218±0.025 && 0.245±0.021 && 0.173±0.020 && 0.100±0.004 &\\
\rule[-1ex]{0pt}{3.5ex}  & & NucleiMix & 0.212±0.018 &3.58& 0.173±0.015 &4.64& 0.225±0.021 &3.37& 0.242±0.019 &-1.36& 0.182±0.018 &5.39& 0.105±0.011 &5.33 \\
\hline

\rule[-1ex]{0pt}{3.5ex} \multirow{8}{6em}{PointNu\\(MoNuSAC: Macrophages)} 

&  \multirow{2}{6em}{epithelial} & Baseline  & 0.266±0.003 && 0.199±0.004 && 0.238±0.003 && 0.290±0.008 && 0.194±0.003 && 0.210±0.004 &\\
\rule[-1ex]{0pt}{3.5ex}  & & NucleiMix & 0.257±0.006 &-3.26& 0.191±0.004 &-3.69& 0.228±0.006 &-3.93& 0.291±0.010 &0.35& 0.186±0.005 &-4.13& 0.211±0.006 &0.16  \\
\cline{2-15}

\rule[-1ex]{0pt}{3.5ex}  &  \multirow{2}{6em}{lymphocyte} & Baseline & 0.390±0.003 && 0.259±0.005 && 0.344±0.004 && 0.483±0.009 && 0.268±0.004 && 0.357±0.009 &\\
\rule[-1ex]{0pt}{3.5ex}  & & NucleiMix & 0.371±0.010 &-4.88& 0.244±0.007 &-5.66& 0.322±0.013 &-6.40& 0.457±0.008 &-5.38& 0.251±0.010 &-6.22& 0.337±0.001 &-5.70 \\
\cline{2-15}

\rule[-1ex]{0pt}{3.5ex}  &  \multirow{2}{6em}{macrophages} & Baseline & 0.144±0.005 && 0.110±0.005 && 0.147±0.007 && 0.192±0.002 && 0.129±0.007 && 0.091±0.003 &\\
\rule[-1ex]{0pt}{3.5ex}  & & NucleiMix & 0.160±0.003 &10.85& 0.126±0.003 &14.50& 0.156±0.003 &6.35& 0.204±0.008 &6.61& 0.136±0.002 &5.43& 0.096±0.003 &5.11 \\
\cline{2-15}

\rule[-1ex]{0pt}{3.5ex}  &  \multirow{2}{6em}{neutrophils} & Baseline & 0.198±0.018 && 0.160±0.012 && 0.214±0.024 && 0.235±0.025 && 0.171±0.016 && 0.112±0.012 &\\
\rule[-1ex]{0pt}{3.5ex}  & & NucleiMix & 0.229±0.005 &16.02& 0.190±0.004 &18.50& 0.250±0.006 &17.00& 0.271±0.009 &15.63& 0.205±0.004 &19.92& 0.141±0.006 &25.52  \\
\hline

\rule[-1ex]{0pt}{3.5ex} \multirow{8}{6em}{SONNET\\(MoNuSAC: Neutrophils)} 

& \multirow{2}{6em}{epithelial} & Baseline  & 0.243±0.005 && 0.177±0.004 && 0.209±0.005 && 0.274±0.016 && 0.168±0.004 && 0.198±0.008 &\\
\rule[-1ex]{0pt}{3.5ex}  & & NucleiMix & 0.246±0.010 &0.96& 0.177±0.008 &0.38& 0.211±0.013 &1.12& 0.278±0.011 &1.34& 0.168±0.011 &0.40& 0.197±0.006 &-0.67  \\
\cline{2-15}

\rule[-1ex]{0pt}{3.5ex}  & \multirow{2}{6em}{lymphocyte} & Baseline & 0.378±0.011 && 0.247±0.008 && 0.328±0.012 && 0.464±0.020 && 0.261±0.011 && 0.351±0.019 &\\
\rule[-1ex]{0pt}{3.5ex}  & & NucleiMix & 0.368±0.009 &-2.82& 0.239±0.004 &-3.50& 0.311±0.010 &-5.18& 0.463±0.024 &-0.36& 0.247±0.008 &-5.61& 0.345±0.015 &-1.71   \\
\cline{2-15}

\rule[-1ex]{0pt}{3.5ex}  & \multirow{2}{6em}{macrophages} & Baseline & 0.167±0.007 && 0.132±0.007 && 0.150±0.008 && 0.192±0.018 && 0.128±0.006 && 0.088±0.006 &\\
\rule[-1ex]{0pt}{3.5ex}  & & NucleiMix & 0.168±0.007 &0.80& 0.130±0.006 &-1.01& 0.147±0.007 &-2.00& 0.208±0.009 &7.97& 0.124±0.005 &-2.61& 0.088±0.007 &-0.38 \\
\cline{2-15}

\rule[-1ex]{0pt}{3.5ex}  & \multirow{2}{6em}{neutrophils} & Baseline & 0.204±0.017 && 0.165±0.016 && 0.214±0.019 && 0.242±0.008 && 0.174±0.017 && 0.104±0.005 &\\
\rule[-1ex]{0pt}{3.5ex}  & & NucleiMix & 0.236±0.023 &15.52& 0.193±0.019 &16.94& 0.250±0.020 &16.98& 0.272±0.023 &12.53& 0.204±0.016 &17.24& 0.114±0.012 &10.29  \\
\hline

\rule[-1ex]{0pt}{3.5ex} \multirow{8}{6em}{PointNu\\(MoNuSAC: Neutrophils)} 

& \multirow{2}{6em}{epithelial} & Baseline  & 0.266±0.003 && 0.199±0.004 && 0.238±0.003 && 0.290±0.008 && 0.194±0.003 && 0.210±0.004 &\\
\rule[-1ex]{0pt}{3.5ex}  &  & NucleiMix & 0.267±0.001 &0.50& 0.198±0.001 &-0.34& 0.234±0.002 &-1.68& 0.296±0.001 &2.07& 0.190±0.002 &-1.89& 0.213±0.001 &1.11  \\
\cline{2-15}

\rule[-1ex]{0pt}{3.5ex}  & \multirow{2}{6em}{lymphocyte} & Baseline & 0.390±0.003 && 0.259±0.005 && 0.344±0.004 && 0.483±0.009 && 0.268±0.004 && 0.357±0.009 &\\
\rule[-1ex]{0pt}{3.5ex}  & & NucleiMix & 0.388±0.012 &-0.51& 0.258±0.010 &-0.51& 0.338±0.013 &-1.84& 0.474±0.018 &-1.86& 0.265±0.012 &-1.00& 0.356±0.017 &-0.37 \\
\cline{2-15}

\rule[-1ex]{0pt}{3.5ex}  & \multirow{2}{6em}{macrophages} & Baseline & 0.144±0.005 && 0.110±0.005 && 0.147±0.007 && 0.192±0.002 && 0.129±0.007 && 0.091±0.003 &\\
\rule[-1ex]{0pt}{3.5ex}  & & NucleiMix & 0.156±0.004 &7.85& 0.123±0.003 &11.78& 0.160±0.003 &8.84& 0.197±0.009 &2.78& 0.138±0.002 &7.24& 0.092±0.002 &0.73 \\
\cline{2-15}

\rule[-1ex]{0pt}{3.5ex}  & \multirow{2}{6em}{neutrophils} & Baseline & 0.198±0.018 && 0.160±0.012 && 0.214±0.024 && 0.235±0.025 && 0.171±0.016 && 0.112±0.012 &\\
\rule[-1ex]{0pt}{3.5ex}  & & NucleiMix & 0.231±0.013 &17.03& 0.187±0.011 &16.63& 0.255±0.017 &19.50& 0.281±0.010 &19.89& 0.205±0.014 &20.31& 0.148±0.008 &32.05 \\
\hline


\rule[-1ex]{0pt}{3.5ex} \multirow{8}{6em}{SONNET\\(MoNuSAC: Macrophages and Neutrophils)} 

& \multirow{2}{6em}{epithelial} & Baseline  & 0.242±0.004 && 0.175±0.004 && 0.209±0.007 && 0.282±0.019 && 0.169±0.006 && 0.203±0.011 &\\
\rule[-1ex]{0pt}{3.5ex}  & & NucleiMix & 0.252±0.006 &3.85& 0.181±0.006 &3.04& 0.212±0.007 &1.27& 0.280±0.011 &-0.71& 0.170±0.006 &0.99& 0.201±0.007 &-1.15  \\
\cline{2-15}

\rule[-1ex]{0pt}{3.5ex}  & \multirow{2}{6em}{lymphocyte} & Baseline & 0.356±0.019 && 0.229±0.013 && 0.306±0.024 && 0.449±0.025 && 0.242±0.020 && 0.338±0.018 &\\
\rule[-1ex]{0pt}{3.5ex}  & & NucleiMix & 0.362±0.010 &1.59& 0.234±0.009 &2.19& 0.316±0.007 &3.38& 0.463±0.018 &3.19& 0.251±0.006 &3.44& 0.348±0.016 &2.86  \\
\cline{2-15}

\rule[-1ex]{0pt}{3.5ex}  & \multirow{2}{6em}{macrophages} & Baseline & 0.167±0.008 && 0.130±0.007 && 0.150±0.008 && 0.194±0.019 && 0.128±0.006 && 0.090±0.008 &\\
\rule[-1ex]{0pt}{3.5ex}  & & NucleiMix & 0.179±0.002 &6.77& 0.139±0.002 &6.65& 0.162±0.002 &8.00& 0.224±0.003 &15.09& 0.136±0.002 &6.79& 0.094±0.002 &4.81 \\
\cline{2-15}

\rule[-1ex]{0pt}{3.5ex}  & \multirow{2}{6em}{neutrophils} & Baseline & 0.205±0.022 && 0.165±0.020 && 0.218±0.025 && 0.245±0.021 && 0.173±0.020 && 0.100±0.004 &\\
\rule[-1ex]{0pt}{3.5ex}  & & NucleiMix & 0.221±0.008 &7.97& 0.180±0.007 &8.87& 0.237±0.008 &9.04& 0.258±0.007 &5.17& 0.194±0.010 &11.95& 0.111±0.006 &11.33\\
\hline

\rule[-1ex]{0pt}{3.5ex} \multirow{8}{6em}{PointNu\\(MoNuSAC: Macrophages and Neutrophils)} 

& \multirow{2}{6em}{epithelial} & Baseline  & 0.266±0.003 && 0.199±0.004 && 0.238±0.003 && 0.290±0.008 && 0.194±0.003 && 0.210±0.004 &\\
\rule[-1ex]{0pt}{3.5ex}  & & NucleiMix & 0.264±0.001 &-0.88& 0.195±0.003 &-1.85& 0.232±0.003 &-2.52& 0.296±0.001 &2.30& 0.189±0.003 &-2.58& 0.215±0.001 &2.06  \\
\cline{2-15}

\rule[-1ex]{0pt}{3.5ex}  & \multirow{2}{6em}{lymphocyte} & Baseline & 0.390±0.003 && 0.259±0.005 && 0.344±0.004 && 0.483±0.009 && 0.268±0.004 && 0.357±0.009 &\\
\rule[-1ex]{0pt}{3.5ex}  & & NucleiMix & 0.386±0.011 &-0.86& 0.258±0.009 &-0.51& 0.340±0.010 &-1.07& 0.465±0.006 &-3.79& 0.266±0.007 &-0.75& 0.339±0.001 &-4.95 \\
\cline{2-15}

\rule[-1ex]{0pt}{3.5ex}  & \multirow{2}{6em}{macrophages} & Baseline & 0.144±0.005 && 0.110±0.005 && 0.147±0.007 && 0.192±0.002 && 0.129±0.007 && 0.091±0.003 &\\
\rule[-1ex]{0pt}{3.5ex}  & & NucleiMix & 0.153±0.002 &6.00& 0.121±0.003 &9.67& 0.157±0.007 &6.80& 0.191±0.001 &-0.52& 0.136±0.006 &5.68& 0.091±0.000 &-0.73  \\
\cline{2-15}

\rule[-1ex]{0pt}{3.5ex}  & \multirow{2}{6em}{neutrophils} & Baseline & 0.198±0.018 && 0.160±0.012 && 0.214±0.024 && 0.235±0.025 && 0.171±0.016 && 0.112±0.012 &\\
\rule[-1ex]{0pt}{3.5ex}  & & NucleiMix & 0.219±0.011 &10.96& 0.178±0.012 &11.23& 0.236±0.010 &10.61& 0.264±0.010 &12.36& 0.192±0.009 &12.50& 0.139±0.009 &23.74 \\
\hline

\end{tabular}
}
\end{center}
\end{table*}

\subsection{Effect of NucleiMix on Class-wise Nuclei Segmentation}
To further assess the impact of nuclei augmentation on segmentation performance, we present class-wise segmentation performance in Table~\ref{tab:seg_class}, where CoNSeP and GLySAC are augmented with miscellaneous nuclei, while MoNuSAC is augmented with macrophages, neutrophils, and both macrophages and neutrophils.

On CoNSeP, the segmentation performance of NucleiMix on miscellaneous nuclei showed consistent gains over the baseline across all evaluation metrics for SONNET, ranging from 8.76\% to 19.61\%. For PointNu, the results were mixed: while the highest gains were obtained using DICE (20.98\%), AJI (18.78\%) and DQ (8.76\%), SQ (-4.86\%) and b-IoU (-4.89\%) were declined. For other nuclei types, both models generally demonstrated performance improvements across all metrics, except for DQ, SQ, PQ, and b-IoU of epithelial nuclei.

On GLySAC, SONNET benefited from NucleiMix, yielding performance gains for all nuclei types over the baseline regardless of evaluation metrics, with the most improvements observed for miscellaneous nuclei, ranging from 1.13\% to 3.89\%. Similarly, PointNu with NucleiMix was generally superior to the baseline across all metrics and nuclei types. 
Miscellaneous nuclei consistently showed improvements, ranging from 0.41\% to 4.66\%. However, slight performance drops were attained for inflammatory in AJI, DQ, and PQ, and for epithelial nuclei in Dice.

On MoNuSAC, using the macrophages augmented dataset, both SONNET and PointNu achieved substantial performance gains for macrophages, ranging from 3.13\% to 11.49\% with SONNET and from 5.11\% to 14.50\% with PointNu. For other types, segmentation performance generally improved, with some exceptions. SONNET exhibited performance drops in SQ and b-IoU for epithelial nuclei and in SQ for neutrophils, while lymphocyte and epithelial nuclei demonstrated performance declines with PointNu. 
By augmenting neutrophils alone, neutrophil achieved the highest improvements regardless of evaluation metrics and segmentation models, ranging from 10.29\% to 17.24\% with SONNET and from 16.63\% to 32.05\% with PointNu. However, the effect of neutrophil augmentation varied across nuclei types and segmentation models. For example, SONNET generally obtained performance gains for epithelial nuclei not for other types. PointNu provided consistent improvements for macrophages, while yielding mixed results for epithelial and lymphocyte nuclei.
As for augmenting both macrophages and neutrophils, the segmentation performance of these two types consistently improved over the baseline across all metrics and segmentation models, except for SQ and b-IoU of macrophages with PointNu. In SONNET, both epithelial and lymphocyte nuclei generally demonstrated improvements except for the SQ and b-IoU of epithelial. In contrast, PointNu showed an overall performance decline for both epithelial and lymphocyte nuclei, except for SQ and b-IoU of epithelial nuclei.

\section{Discussion}

Data augmentation techniques are widely used to address data imbalance and to overcome the limitations of small datasets. Mix-based data augmentation methods typically blend parts from existing images to generate composite images. In the context of nuclei instance segmentation, such methods can be adopted to increase the representation of rare-type nuclei. For instance, CutMix and CopyPaste randomly select and paste nuclei to artificially increase the number of rare-type nuclei. GradMix, on the other hand, adopts a more targeted approach, selecting pairs of major-type and rare-type nuclei. It requires the major-type nucleus to be larger, allowing the rare-type nucleus to be placed within its region, and the neighboring pixels of the rare-type nucleus are updated using a weighted sum of their original surroundings. While these methods can integrate the existing rare-type nuclei into new environments, none of them analyzes or leverages the relationship between the original neighboring environments of rare-type nuclei and the new environments into which where they are inserted. In contrast, NucleiMix improves upon these methods by adopting a context-aware probabilistic sampling strategy, which probabilistically selects appropriate positions for augmenting rare-type nuclei, under the assumption that the more similar environments between the original and target positions, the more realistic the augmentation will be. 

Diffusion Models have achieved unprecedented success in solving inverse problems, and we leverage this capability to mitigate artifacts introduced by data augmentations in nuclei instance segmentation. Diffusion models can be adopted to generate new instances of nuclei as a sufficient number of instances are provided. When it comes to rare-type nuclei, it becomes impossible to reliably train any model to produce realistic nuclei. Pre-trained diffusion models can also be used for direct inpainting of surrounding regions as mixing existing nuclei with new environments; however, this often results in obscure boundaries and inconsistent surroundings. To address this, NucleiMix introduces a progressive inpainting strategy that imposes spatial restrictions on the inpainted areas, ensuring that the newly generated environments are coherent while maintaining sharp, well-defined boundaries for rare-type nuclei (Fig. \ref{fig:procedure}). This approach, however, comes at the cost of increased computational complexity per instance. Further investigation is needed to expedite the inpainting process and reduce computational costs.

NucleiMix demonstrated improvements in both segmentation and classification performance for various types of nuclei in three public datasets. However, other augmentation methods often resulted in worse performance than the baseline where no synthesized nuclei were added and used. The improved performance obtained by NucleiMix may be ascribable to its context-aware probabilistic sampling for candidate location selection and progressive inpainting for realistic reconstruction of new environments. The former ensures that rare-type nuclei are placed in similar environments, and the latter seamlessly integrates the rare-type nuclei into their new environment. In contrast, other augmentation methods tend to produce unrealistic or awkward nuclei, which could adversarially affect the performance of nuclei instance segmentation. Moreover, NucleiMix generally improved the nuclei instance segmentation performance with varying numbers of augmented rare-type nuclei. Overall, k=600 was found to be the optimal number of augmented nuclei. The smaller numbers of augmented nuclei (e.g., k=200) resulted in fewer improvements, and adding more augmented nuclei (e.g., k=800) did not consistently lead to better performance. This suggests that the performance tends to saturate after a certain number of augmented nuclei is introduced.

This study has several limitations.
First, we adopted GMM to learn the distribution of rare-type nuclei in the feature space and to predict the likelihood of major-type and background locations. Although the nuclei instance segmentation performance was improved, the accuracy of the predicted candidate locations cannot be fully guaranteed and may vary on a case-by-case basis. 
Second, the use of a diffusion model for inpainting introduces stochasticity and leads to time-consuming reconstructions, which are its inherent limitations. Future work will focus on leveraging advanced samplers designed to accelerate the inference speed of the diffusion model.
Third, NucleiMix affected the nuclei instance segmentation performance not only for rare-type nuclei but also for major-type nuclei. This may be attributed to the Copy-Replace operation applied to major-type nuclei, which reduces their overall number. The relationship between the major-type and rare-type nuclei with respect to nuclei instance segmentation, as well as the impact of the Copy-Replace operation, needs further investigation.
Fourth, the performance gains appeared to plateau at higher augmentation levels. This may be due to the reduced variability resulting from repeatedly sampling semantically similar rare-major and rare-background pairs. Although NucleiMix introduces realistic nuclei, its effectiveness is ultimately constrained by the diversity of available rare-type nuclei for augmentation. Future work will explore how to increase global diversity and meanwhile maintaining contextual consistency of the sampled pairs. For example, we plan to track the augmented regions and produce a spatial augmentation heatmap, which can be used to penalize areas that have already used for augmentation, thereby encouraging more spatially diverse placements.
Fifth, the effect of NucleiMix varied across different nuclei types. While the segmentation performance of the augmented nuclei generally improved, the performance of other types showed mixed results. This phenomenon can be attributed to the interdependent learning dynamics of deep learning models, where enhancing the representation of one class can shift feature boundaries in a way that inadvertently affects others. Advanced strategies to mitigate such adverse effects on non-augmented types may further enhance the overall effectiveness of NucleiMix.
Sixth, we have not explicitly analyzed the surroundings of nuclei whether they are similar or not. However, it is generally accepted in histopathology that nuclei (or cells) of the same type tend to reside in similar tissue environments. For instance, epithelial cells lining a gland typically share a similar environment, characterized by a lumen surrounded by epithelial cells, which are encompassed by stromal cells. Nevertheless, depending on the tissue structure and disease status, the specific environment may vary. In particular, the miscellaneous category contains multiple nucleus types and unspecified types, which could possess different surroundings. Such differences may affect the proposed method, thereby affecting the performance of nuclei synthesis.
Last, three public datasets and two backbone models were utilized in this study. To further confirm the effectiveness of NucleiMix, an extended validation study including additional datasets and backbone models should be followed.

\section{Conclusion}
We propose NucleiMix, a data augmentation method that enhances the performance and robustness of nuclei segmentation and classification. NucleiMix samples semantically similar positions for augmenting rare-type nuclei and employs a diffusion-based progressive inpainting strategy for realistic integration. Superior results on three public datasets (CoNSeP, GLySAC, and MoNuSAC) with two backbone models (SONNET and PointNu) demonstrate its effectiveness in addressing data imbalance. Future work will explore broader applications of NucleiMix across various tissue and disease types.

\section*{Acknowledgments}
This study was supported by a grant of the National Research Foundation of Korea (NRF) (No. 2021R1A2C2014557 and No. RS-2024-00397293) and by the Ministry of Trade, Industry and Energy(MOTIE) and Korea Institute for Advancement of Technology (KIAT) through the International Cooperative R\&D program (P0022543).

\bibliographystyle{model2-names.bst}
\bibliography{refs}

\end{document}